\documentclass{article}
\pdfoutput=1
\RequirePackage[OT1]{fontenc}
\RequirePackage{amsmath,amsfonts,amssymb,amsthm}
\RequirePackage{natbib}
\usepackage{authblk}
\RequirePackage[colorlinks,citecolor=blue,urlcolor=blue]{hyperref}

\usepackage[toc,page]{appendix}

\usepackage[normalem]{ulem}
\usepackage{graphicx,bookmark}
\usepackage{booktabs,mathtools}
\usepackage{rotating}
\usepackage{here}
\usepackage[margin=1.25in]{geometry}
\usepackage{dsfont}
\usepackage{booktabs}
\usepackage{multirow}
\numberwithin{equation}{section}

\newtheorem{theorem}{Theorem}[section]
\newtheorem{lemma}[theorem]{Lemma}
\theoremstyle{remark}

\newtheorem{proposition}{Proposition}
\newtheorem{remark}[proposition]{Remark}

\newcommand{\eps}{\varepsilon}

\newcommand{\bPsi}{\boldsymbol{\Psi}}
\DeclareMathOperator{\sLM}{sLM}
\DeclareMathOperator{\sLMg}{sLM^{\star}}
\DeclareMathOperator{\sLMh}{sLM_{\textsc{hr}}}
\DeclareMathOperator{\sLMgh}{sLM^{\star}_{\textsc{hr}}}

\begin{document}
\title{Testing for threshold effects in the TARMA framework}

\author[2]{Greta Goracci}
\author[2]{Simone Giannerini}
\author[1]{Kung-Sik Chan}
\author[3,4,5]{Howell Tong}

\affil[1]{\small Department of Statistics and Actuarial Science,
University of Iowa, Iowa City, USA}
\affil[2]{Department of Statistical Sciences, University of Bologna, Italy}
\affil[3]{School of Mathematical Science, University of Electronic Science and Technology, Chengdu China}
\affil[4]{Center for Statistical Science, Tsinghua University, China}
\affil[5]{London School of Economics and Political Science, U.K.}

\date{\small \today}

\maketitle


\begin{abstract}
We present supremum Lagrange Multiplier tests to compare a linear ARMA specification against its threshold ARMA extension. We derive the asymptotic distribution of the test statistics both under the null hypothesis and contiguous local alternatives. Moreover, we prove the consistency of the tests. The Monte Carlo study shows that the tests enjoy good finite-sample properties, are robust against model mis-specification and their performance is not affected if the order of the model is unknown. The tests present a low computational burden and do not suffer from some of the drawbacks that affect the quasi-likelihood ratio setting. Lastly, we apply our tests to a time series of standardized tree-ring growth indexes and this can lead to new research in climate studies.
\end{abstract}

%
%
%

\section{Introduction}

Threshold autoregressive models have gained popularity in Economics, Biology, and many other fields, \citep{Ton90,Cha17b,Cha09,Ton11,Han11}. In particular, TARMA models, introduced in \cite{Ton78} and \cite{Ton80}, are non-linear models with a regime-switching mechanism specifying an ARMA sub-model in each regime. They include two particular models of independent interest: the threshold autoregressive (TAR) model and the threshold moving-average (TMA) model. By combining both the TAR model and the TMA model, TARMA models are parsimonious and yet rich models for non-linear time series analysis (see e.g. \citet{Gor20,Gor21}).  \cite{Li11b} developed the theory for least squares estimation of parameters of the general TARMA$(p,q)$ model by assuming stationarity and ergodicity. Nevertheless, the conditions for ergodicity derived by \cite{Lin99} are quite restrictive. A full characterization of the long-run probabilistic behaviour of TARMA models was not available until \citet{Cha19}, which derived the necessary and sufficient conditions for the ergodicity of the first-order TARMA model. Moreover, they provided a complete parametric classification of the first-order TARMA into regions where it is (geometrically) ergodic, null recurrent and transient.
\par \noindent
Considerable efforts have been produced to test whether a threshold model provides a better fit with respect to its linear counterpart. Most contributions focus on AR-type models. For instance, \citet{Pet86} developed a portmanteau test based on cumulative sums of standardized residuals from an autoregressive fit. \citet{Tsa98} studied a variation of such test. \citet{Luu88} proposed a Lagrange Multiplier test for linearity against a large class of non-linear models that includes the TAR specification. A Lagrange Multiplier test was also developed in \citet{Won97, Won00} for TAR models with conditional heteroscedasticity. Quasi-likelihood ratio tests were studied in \citet{Cha90a,Cha90b,Cha91} up to the recent test for threshold diffusion of \cite{Su17}. For a review see also \cite{Ton11}.
\par
The framework of threshold models that includes a moving-average component has been under-investigated probably due to the mathematical difficulties that arise when the moving-average component is incorporated in a non-linear setting. However, since data are almost always affected by measurement error, the threshold ARMA framework is more appropriate than a pure autoregressive approach. Indeed, it is known that a AR process plus measurement error becomes a ARMA process. Likewise, it can be proved that a TAR process of order $p$ corrupted with additive measurement noise may be approximated by a TARMA model of order $(p, p)$. The adoption of the TARMA framework is not a minor point since a high autoregressive order may be needed to approximate the moving-average component, at the expense of loss in power of the test. In the framework of MA-type models, \citet{Lin05} investigated a quasi-likelihood ratio test for the MA model against its threshold extension. They proved that, under the null hypothesis, the test statistic converges weakly to a functional of the centered Gaussian process. Their results were extended in \citet{Li08} to the case with GARCH errors. More recently, \citet{Li11} developed a quasi-likelihood ratio statistic to test the presence of thresholds in ARMA processes. They use a stochastic permutation device to build the distribution of the statistic under the null hypothesis.
\par
In this paper we extend the work of \citet{Cha90a} and \citet{Lin05} and propose supremum Lagrange multiplier test statistics (supLM) to determine whether a TARMA model fits a stationary time series significantly better than an ARMA model. One of the main advantages of using a Lagrange Multiplier approach over likelihood-ratio tests is that it does not need estimating the model under the alternative hypothesis. We prove that both under the null hypothesis and contiguous local alternatives the asymptotic distribution of the test statistics reduces to the same functional of a Gaussian process that is centered under the null and non centered under the alternative. The results extend the work of \cite{Lin05} on the weak convergence of linear marked empirical processes with infinitely many markers to the case where the underlying process is an ARMA$(p,q)$. Moreover, we prove the consistency of our tests and show that they have non-trivial power against local alternatives.
\par
In order to test the ARMA$(p,q)$ specification against its TARMA extension we propose two supLM statistics: in the first one, denoted by $\sLM$, only the autoregressive part is tested for threshold non-linearity whereas in the second statistic, denoted by $\sLMg$, both the autoregressive and the moving-average part are tested. As it will be clear, the two statistics are different; in particular, the $\sLMg$ statistic does not reduce to the $\sLM$ when the moving-average part is either absent or does not change across regimes. This is reflected on the different finite sample behaviour of the tests.
  \par
 We explore systematically the performance of our supLM tests and compare them with the quasi-likelihood ratio test of \citet{Li11} (qLR from now on): the extensive simulation study shows clearly that our tests have better size and power while enjoying a much lower computational burden.
 Furthermore, the two supLM tests are robust against model mis-specification and the performance of the tests is not adversely affected if the order of the ARMA process is unknown and is selected through the Hannan-Rissanen procedure. Lastly, we apply our test to the time series of standardized tree-ring growth indexes. We show that the TARMA(1,1) specification can provide a better fit with respect to the accepted ARMA model. We believe that the TARMA framework can lead to a better understanding of the tree-ring dynamics and lead to novel directions of research where the econometric approach is properly adopted in climate studies.
\par \noindent
The rest of the paper is organized as follows: in Section~\ref{sec:notation} we present our setting and the tests; in Section~\ref{sec:null} we derive the distributions under the null hypothesis and tabulate the empirical quantiles. In Section~\ref{sec:alternatives} we derive the asymptotic distribution of the statistics under local contiguous alternatives and prove consistency of the tests. Section~\ref{sec:finite_performance} contains a Monte Carlo study to assess the finite sample performance of our proposals. We also investigate the behaviour of the tests under model mis-specification and when the order of the tested model is unknown. In Section~\ref{sec:real} we apply our tests to a tree-ring time series whereas some discussion and the conclusions are reported in Section~\ref{sec:conclusions}. All the proofs are detailed in the Supplementary Material, that also contains additional results regarding both the simulation study and the tree-ring data analysis.
\section{Notation and preliminaries}\label{sec:notation}
Let the time series $\{X_t:t=0,\pm 1,\pm2,\dots\}$ follow the threshold autoregressive moving-average model defined by the difference equation:
\begin{align}\label{eq:TARMA}
  X_t&=\phi_{10}+\sum_{k=1}^{p}\phi_{1k}X_{t-k}+\eps_t-\sum_{s=1}^{q}\theta_{1s} \eps_{t-s}\nonumber\\
     &+\left(\Psi_{10}+\sum_{k=1}^{p}\Psi_{1k}X_{t-k}-\sum_{s=1}^{q}\Psi_{2s} \eps_{t-s}\right)I(X_{t-d}\leq r).
\end{align}
\noindent
In the case where the moving-average parameters are fixed across regimes, the sum involving $\Psi_{2s}$, $s=1,\dots,q$ is absent.
The innovations $\{\eps_t\}$ are independent and identically distributed (iid) random variables such that, for each $t$, $\eps_t$ has zero mean, finite variance $\sigma^2$ and is independent of $X_{t-1}$, $X_{t-2}$, \dots . {Note that the iid assumption can be relaxed to a stationary ergodic martingale difference sequence with respect to the $\sigma$-algebra generating by $\eps_s$ with $s<t$. In such a case the proofs do not change.} The positive integers $p$ and $q$ are the autoregressive and moving-average orders, respectively; $d$ is the delay parameter that takes positive integer values. We assume $p,q,d$ to be known. Moreover, $I(\cdot)$ is the indicator function and $r\in\mathds{R}$ is the threshold parameter. For notational convenience, we abbreviate $I(X_t\leq r)$ by $I_r(X_t)$. All the results are derived conditionally upon the $p$ initial values of $\{X_t\}$. Let
\begin{align}\label{eq:pars}
\boldsymbol\phi=\left(\phi_{10},\phi_{11},\dots,\phi_{1p}\right)^\intercal; \quad \boldsymbol\theta=\left(\theta_{11},\dots,\theta_{1q}\right)^\intercal;\quad
\end{align}
and $\bPsi$ be the vector containing the parameters that are tested. The true parameters of the model are
$
\boldsymbol\eta=\left(\boldsymbol{\zeta}^\intercal,\sigma^2,\boldsymbol\Psi^\intercal\right)^\intercal
$,
where
\begin{align}\label{eq:pars2}
\boldsymbol{\zeta} =(\boldsymbol\phi^\intercal,\boldsymbol\theta^\intercal)^\intercal, \bPsi = \left(\Psi_{10},\Psi_{11},\dots,\Psi_{1p}\right)^\intercal,  \text{ if only the AR part changes,}\\
\boldsymbol{\zeta}= \boldsymbol\phi,
\bPsi= \left(\Psi_{10},\Psi_{11},\dots,\Psi_{1p},\Psi_{21},\dots,\Psi_{2q}\right)^\intercal,  \text{ if both the AR and MA parts change.}
\end{align}
We test whether a threshold ARMA$(p,q)$ model provides a significantly better fit than the linear ARMA$(p,q)$ model. To this end, we develop two Lagrange multiplier test statistics for the hypothesis
$$
\begin{cases}
  H_0&:\boldsymbol\Psi=\boldsymbol 0 \\
  H_1&: \boldsymbol\Psi\neq\boldsymbol 0,
\end{cases}
$$
where $\boldsymbol 0$ is the vector with all zeroes. The statistic for testing the threshold effect in the AR parameters is denoted as $\sLM$, whereas $\sLMg$ is the statistic for the general test where both the AR and the MA parameters change across regimes. Under $H_0$ the process follows a linear ARMA$(p,q)$ model:
\begin{equation}\label{eqn:ARMA1}
  X_t=\phi_{10}+\sum_{k=1}^{p}\phi_{1k}X_{t-k}+\eps_t-\sum_{s=1}^{q}\theta_{1s}\eps_{t-s}.
\end{equation}
To derive the asymptotic features of the test, we assume the model to be ergodic and invertible both under the null and the alternative hypothesis. Let the autoregressive and moving-average polynomials be defined as follows:
\begin{align*}
\phi(z)=1-\phi_{11}z-\phi_{12}z^2-\dots-\phi_{1p}z^p; \quad \theta(z)=1-\theta_{11}z-\theta_{12}z^2-\dots-\theta_{1q}z^q.
\end{align*}
  Assumption A1 below ensures the ergodicity and invertibility of model (\ref{eqn:ARMA1}) and also avoids certain degeneracy. For more details see \citet{Cha10} and \citet{Cha19}.
\begin{description}
  \item[A1.]  $\phi(z)\neq0$ and $\theta(z)\neq 0$ for all $z\in\mathds{C}$ such that $|z|\leq 1$ and they do not share common roots.
\end{description}
\par
From now on we fully develop the theory for the general statistic $\sLMg$. Unless otherwise specified, the results hold also for the statistic $\sLM$. Suppose we observe $X_1,\dots,X_n$.
We develop the Lagrange multiplier test based on the Gaussian likelihood conditional on the initial values $X_0,X_{-1},\ldots,X_{-p+1}$:
\begin{equation}\label{log-like}
\ell = -\frac{n}{2}\log (\sigma^2 2\pi) -\frac{1}{2\sigma^2}\sum_{t=1}^n \eps_t^2,
\end{equation}
where, by an abuse of notation, we set
\begin{align}\label{eq:eps}
\eps_t=X_t&-\left\{\phi_{10}+\sum_{k=1}^{p}\phi_{1k}X_{t-k}-\sum_{s=1}^{q}\theta_{1s}\eps_{t-s}\right\} \nonumber\\
&-\left\{\Psi_{10}+\sum_{k=1}^{p}\Psi_{1k}X_{t-k}-\sum_{s=1}^{q}\Psi_{2s}\eps_{t-s}\right\}I_r\left(X_{t-d}\right)
\end{align}
and $\eps_0,\eps_{-1},\ldots,\eps_{-q+1}$ are set to be zero. Clearly, $\eps_t$ is a function of $\boldsymbol\eta$ and $r$, but we omit the arguments for simplicity. As in Eq~(\ref{eq:TARMA}) for the $\sLM$ test, the sum involving $\Psi_{2s}$, $s=1,\dots,q$ is absent.
\par\noindent
{Let $\partial\ell/\partial\boldsymbol\eta$ be the score vector, whose components are
\begin{align}
&\frac{\partial\ell}{\partial\eta_i}=\sum_{t=1}^{n}\frac{\eps_t^2-\sigma^2}{2\sigma^4}, \qquad &\text{ if $\eta_i=\sigma^2$,}\\
&\frac{\partial\ell}{\partial\eta_i}=-\sum_{t=1}^{n}\frac{\eps_t}{\sigma^2} \frac{\partial\eps_t}{\partial\eta_i}, \qquad &\text{ otherwise.}
\end{align}
and $\partial\ell/\partial\bPsi$ be the derivatives of the log-likelihood with respect to $\bPsi$.} Moreover, let
\begin{align*}
\frac{\partial\eps_t}{\partial\boldsymbol\phi}&=\left(\frac{\partial\eps_t}{\partial\phi_{10}},\frac{\partial\eps_t}{\partial\phi_{11}},\dots,\frac{\partial\eps_t}{\partial\phi_{1p}}\right)^\intercal;
\frac{\partial\eps_t}{\partial\boldsymbol\theta}=\left(\frac{\partial\eps_t}{\partial\theta_{11}},\dots,\frac{\partial\eps_t}{\partial\theta_{1q}}\right)^\intercal;\\
\end{align*}
and
\begin{equation*}
\frac{\partial\eps_t}{\partial\boldsymbol\Psi} =
\begin{cases}
\left(\frac{\partial\eps_t}{\partial\Psi_{10}},\frac{\partial\eps_t}{\partial\Psi_{11}},\dots,\frac{\partial\eps_t}{\partial\Psi_{1p}},\frac{\partial\eps_t}{\partial\Psi_{21}},\dots,\frac{\partial\eps_t}{\partial\Psi_{2q}}\right)^\intercal, & \text{for the $\sLMg$ statistic}  \\
  \left(\frac{\partial\eps_t}{\partial\Psi_{10}},\frac{\partial\eps_t}{\partial\Psi_{11}},\dots,\frac{\partial\eps_t}{\partial\Psi_{1p}}\right)^\intercal, & \text{for the $\sLM$ statistic}.
\end{cases}
\end{equation*}

Finally
\begin{equation}\label{eq:Fmatrix}
\mathcal{I}_n(r)
      =\begin{pmatrix}
                  \mathcal{I}_{n,11}    & \mathcal{I}_{n,12}(r) \\
                  \mathcal{I}_{n,12}^\intercal(r) & \mathcal{I}_{n,22}(r)
       \end{pmatrix}
  =\begin{pmatrix}
              \frac{\partial^2\ell}{\partial\boldsymbol{\zeta}\partial\boldsymbol{\zeta}^\intercal}   & \frac{\partial^2\ell}{\partial\boldsymbol{\zeta}\partial\boldsymbol{\Psi}^\intercal} \\
              \frac{\partial^2\ell}{\partial\boldsymbol{\Psi}\partial\boldsymbol{\zeta}^\intercal} &
\frac{\partial^2\ell}{\partial\boldsymbol{\Psi}\partial\boldsymbol{\Psi}^\intercal}
   \end{pmatrix}.
\end{equation}
 The TARMA$(p,q)$ model under the null hypothesis can be estimated by using the method of the maximum likelihood.
Let $\frac{\partial \hat{\ell} }{\partial \boldsymbol\Psi}(r)$ and $\hat{\mathcal{I}}_{n}(r)$ be equal to $\frac{\partial \ell }{\partial \boldsymbol\Psi}(r)$ and  $\mathcal{I}_n(r)$, respectively, evaluated at the maximum likelihood estimates for the ARMA part and with $\bPsi=\boldsymbol 0$.
\par
Under the null hypothesis, the threshold parameter $r$ is absent thereby the standard asymptotic theory is not applicable. To cope with this issue, we firstly develop the Lagrange multiplier test statistic as a function of $r$ ranging in a set $\mathcal{R}$. Then, for all the values $r \in \mathcal{R}$, we compute the test statistic and, finally, we take the overall test statistic as the supremum on $\mathcal{R}$. We set $\mathcal{R}=[r_L,r_U]$, $r_L$ and $r_U$ being some percentiles of the data. This approach has become widely used in the literature of tests involving threshold models, see, for instance, \citet{Cha90a}, \citet{Lin05}, \citet{Li11} and \citet{Cha20}.
Our test statistic is
  \begin{align}
  T_n&=\sup_{r\in[r_L,r_U]}T_{n}(r);\label{eq:Tn}\\
 T_{n}(r)&=\left(\frac{\partial \hat{\ell} }{\partial \boldsymbol\Psi}(r)\right)^\intercal \left(\hat{\mathcal{I}}_{n,22}(r)-\hat{\mathcal{I}}_{n,21}(r)\hat{\mathcal{I}}_{n,11}^{-1} \hat{\mathcal{I}}_{n,12}(r)\right)^{-1}\frac{\partial \hat{\ell} }{\partial \boldsymbol\Psi}(r).\label{eq:Tnr}
  \end{align}
\section{The null distribution}\label{sec:null}
In this section we derive the asymptotic distribution of $T_n$ under the null hypothesis that $\{X_t\}$ follows an ARMA$(p,q)$ process. Unless stated otherwise, all the expectations are taken under the true probability distribution for which $H_0$ holds. Also, $o_p(1)$ denotes the convergence in probability to zero as $n$ increases and $\|\cdot\|$ is the $L^2$ matrix norm (the Frobenius' norm, i.e. $\|A\|=\sqrt{\sum_{i=1}^{n}\sum_{j=1}^{m}|a_{ij}|^2}$, where $A$ is a $n\times m$ matrix). Moreover, let $\mathcal{D}_\mathds{R}(a,b)$, $a<b$ be the space of functions from $(a,b)$ to $\mathds{R}$ that are right continuous with left-hand limits. $\mathcal{D}_\mathds{R}(a,b)$ is equipped with the topology of uniform convergence on compact sets, see \citet{Bil68} for more details.
\par\noindent
In the following two lemmas, we rewrite $\partial\eps_t/\partial\boldsymbol\phi$, $\partial\eps_t/\partial\boldsymbol\theta$ and  $\partial\eps_t/\partial\boldsymbol\Psi$ as functions of the roots of the characteristic moving-average polynomial $\theta(\cdot)$ and provide a uniform approximation of the matrix $\mathcal{I}_n(r)$.
\begin{lemma}\label{lemma:score}
  Under Assumption A1 and under $H_0$, the following holds. Let $\alpha_j$, $j=0,1,2,\dots$, satisfy the difference equation    $\alpha_0=1$, $\alpha_j-\sum_{s=1}^{q}\theta_s\alpha_{j-s}=0$, whose initial conditions are $\alpha_j=0$, for $j< 0$. Then
    \begin{enumerate}
    \item the $(k+1)$-th entry of $\partial\eps_t/\partial\boldsymbol\phi$ is
   \[
     -\sum_{j=0}^{t-1}\alpha_j, \quad \text{ if } k=0; \qquad
     -\sum_{j=0}^{t-k}\alpha_jX_{t-k-j}, \quad \text{ if } 1\leq k\leq p.
   \]
    \item the $k$-th entry of $\partial\eps_t/\partial\boldsymbol\theta$ is
    \[
    \sum_{j=0}^{t-k}\alpha_j\eps_{t-k-j}.
    \]
    \item the $(k+1)$-th entry of $\partial\eps_t/\partial\boldsymbol\Psi$ is
    \begin{align*}
     -\sum_{j=0}^{t-1}\alpha_jI_r(X_{t-d-j}), \;& \text{ if } k=0; \quad
     -\sum_{j=0}^{t-k}\alpha_jX_{t-k-j}I_r(X_{t-d-j}),  \;\text{ if } 1\leq k\leq p; \\
     \sum_{j=0}^{t-k}\alpha_j\eps_{t-k-j}I_r(X_{t-d-j}), \;& \text{ if } p+1\leq k\leq p+q+1
   \end{align*}
where, in the preceding equation, the components corresponding to $p+1\leq k\leq p+q+1$ are absent in the $\sLM$ test.
    \end{enumerate}
    \end{lemma}
  \begin{remark}
 The Lemma~(\ref{lemma:score}) is an alternative way of representing the derivatives that may simplify the derivations. Under the null hypothesis, we obtain the same expansion for the partial derivatives of $\eps_t$ described by Eq (6.3) and (6.4) in  \cite{Lin05}.
  \end{remark}
  \begin{lemma}\label{lemma:Fisher}
  Under Assumption A1 and under $H_0$, as $n\to\infty$ it holds that
\begin{equation}\label{eq:FMapprox}
\mathcal{I}_n(r)
  =\begin{pmatrix}
\sum_{t=1}^{n}\frac{1}{\sigma^2}\left(\frac{\partial\eps_t}{\partial\boldsymbol{\zeta}}\right) \left(\frac{\partial\eps_t}{\partial\boldsymbol{\zeta}}\right)^\intercal &               \sum_{t=1}^{n}\frac{1}{\sigma^2}\left(\frac{\partial\eps_t}{\partial\boldsymbol{\zeta}}\right) \left(\frac{\partial\eps_t}{\partial\boldsymbol{\Psi}}\right)^\intercal  \\
\sum_{t=1}^{n}\frac{1}{\sigma^2}\left(\frac{\partial\eps_t}{\partial\boldsymbol{\Psi}}\right) \left(\frac{\partial\eps_t}{\partial\boldsymbol{\zeta}}\right)^\intercal  & \sum_{t=1}^{n}\frac{1}{\sigma^2}\left(\frac{\partial\eps_t}{\partial\boldsymbol{\Psi}}\right) \left(\frac{\partial\eps_t}{\partial\boldsymbol{\Psi}}\right)^\intercal    \end{pmatrix} + o_p(n).
\end{equation}
\end{lemma}
Now, define $ \nabla_n(r)=\left(\nabla^\intercal_{n,1},\nabla^\intercal_{n,2}(r)\right)^\intercal$, where
 \begin{align}\label{eq:nabla}
 \nabla_{n,1}=\left(\frac{1}{\sqrt{n}}\sum_{t=1}^{n}\frac{\eps_t}{\sigma^2} \left(\frac{\partial\eps_t}{\partial\boldsymbol\zeta}\right)^\intercal\right)^\intercal;
 \nabla_{n,2}(r)=\left(\frac{1}{\sqrt{n}}\sum_{t=1}^{n}\frac{\eps_t}{\sigma^2} \left(\frac{\partial\eps_t}{\partial\boldsymbol\Psi}\right)^\intercal\right)^\intercal.
 \end{align}
Moreover, define
\begin{align*}
\frac{1}{\sigma^2}E\left[D(r)D(r)^\intercal\right]=\Lambda(r)=\left(\begin{array}{lr}
                                                                         \Lambda_{11} & \Lambda_{12}(r) \\
                                                            \Lambda_{12}^\intercal(r) & \Lambda_{22}(r)
                                                                        \end{array}\right),
\end{align*}
where the $\Lambda_{ij}(r)$ have same dimension as the $\mathcal{I}_{n,ij}(r)$ of Eq.~(\ref{eq:Fmatrix}) and
\begin{align*}
D(r)=&\left(\sum_{j=0}^{\infty}\alpha_j,\sum_{j=0}^{\infty}\alpha_jX_{t-1-j},\ldots,\sum_{j=0}^{\infty}\alpha_jX_{t-p-j},
\sum_{j=0}^{\infty}\alpha_j\eps_{t-1-j},\ldots,\sum_{j=0}^{\infty}\alpha_j\eps_{t-q-j},\right.\\
&\phantom{\left(\right.}\left.\sum_{j=0}^{\infty}\alpha_jI_r(X_{t-d-j}),\sum_{j=0}^{\infty}\alpha_jI_r(X_{t-d-j})X_{t-1-j},\ldots,\sum_{j=0}^{\infty}\alpha_jI_r(X_{t-d-j})X_{t-p-j}\right.\\
&\phantom{\left(\right.}\left.\sum_{j=0}^{\infty}\alpha_jI_r(X_{t-d-j})\eps_{t-1-j},\ldots,\sum_{j=0}^{\infty}\alpha_jI_r(X_{t-d-j})\eps_{t-q-j}\right)^\intercal
\end{align*}
with the $\alpha$'s defined as in Lemma~ \ref{lemma:score} and where, for the $\sLM$ statistic, the components from position $p+2$ to $p+q+2$ and the last $q$ of $D(r)$ are absent.
\par
In the following Lemma, we show some properties of the $\nabla$'s and $\Lambda$'s under the null hypothesis and Assumption A1.
\begin{proposition}\label{prop:asymptotics}
Under Assumption A1 and under $H_0$, we have the following:
\begin{enumerate}
\item For each $\boldsymbol{\eta}$, the matrix $\Lambda(r)$ is positive definite.
\item
\begin{align*} \sup_{r\in[a,b]} & \left\|\left(\frac{\mathcal{I}_{n,22}(r)}{n} - \frac{\mathcal{I}_{n,21}(r)}{n}\left(\frac{\mathcal{I}_{n,11}}{n}\right)^{-1}\frac{\mathcal{I}_{n,12}(r)}{n}\right)^{-1} - \right.\\
& \left.\left(\Lambda_{22}(r)-\Lambda_{21}(r)\Lambda_{11}^{-1}\Lambda_{12}(r)\right)^{-1}\right\| = o_p(1).
\end{align*}
\item \[ \sup_{r\in[a,b]}\left\|\frac{1}{\sqrt{n}}\frac{\partial\hat\ell}{\partial\boldsymbol\Psi}(r)- \left(\nabla_{n,2}(r)-\Lambda_{21}(r)\Lambda_{11}^{-1}\nabla_{n,1}\right)\right\|
=o_p(1).
\]
\end{enumerate}
\end{proposition}
Note that in the above proposition, $\left(\nabla_{n,2}(r)-\Lambda_{21}(r)\Lambda_{11}^{-1}\nabla_{n,1}\right)$ is a linear marked empirical process with infinitely many markers. Similarly to \cite{Lin05,Li11}, we rely on Assumption A2:
 \begin{description}
   \item[A2.] $\eps_t$ has a continuous and strictly positive density on the real line and $E\left[\eps^4_t\right]$ is finite.
 \end{description}
In the following theorem we derive the null asymptotic distribution of the Lagrange Multiplier test statistic $T_n$.
 \begin{theorem}\label{th:null}
   Let  $\left\{\xi(r),\;r\in\mathds{R}\right\}$ be a centered Gaussian process with covariance kernel  $$\Sigma(r_1,r_2)=\Lambda_{22}(r_1\wedge r_2)-\Lambda_{21}(r_1)\Lambda_{11}^{-1}\Lambda_{12}(r_2).$$
   Then, under $H_0$ and Assumptions A1 and A2, $T_n$ converges weakly to
   $$\sup_{r\in[r_L,r_U]}\xi(r)^\intercal\left(\Lambda_{22}(r)-\Lambda_{21}(r)\Lambda_{11}^{-1}\Lambda_{12}(r)\right)^{-1}\xi(r).$$

 \end{theorem}

\subsection{Empirical quantiles of the null distribution}

In Table~\ref{tab:q1} we tabulate the empirical quantiles of the null asymptotic distribution of our supLM statistics at levels 90\%, 95\%, 99\% and 99.9\% for autoregressive orders from 1 to 4 and moving-average orders from 1 to 2. The threshold is searched between the 25th and the 75th percentiles of the sample distribution. For each order, the results have been obtained from 10000 simulated series of length 1000 and are presented in Table~\ref{tab:q1}. The quantiles of the asymptotic distribution of the $\sLM$ statistic do not depend upon the moving-average parameters and are in good agreement with those of \cite{Cha91}, table~1, that refer to testing the AR against the TAR model (In that case, the length of the series was 200 and the number of replications 1000). The rightmost part of Table~\ref{tab:q1} contains the quantiles for the $\sLMg$ statistic. Indeed, even when the moving-average parameters do not change across regimes, the $\sLMg$ statistic does not reduce to the $\sLM$ statistic. Furthermore, the asymptotic distribution of the $\sLM$ statistic is equivalent to that of \cite{Cha91} since the vector $\nabla_{n}(r)$ of Eq.~(\ref{eq:nabla}) does not contain the partial derivatives w.r.t. to the moving-average part. Notably, the asymptotic behaviour of the two statistics depends only upon the dimension of the parameter vector $\bPsi$, irrespectively of its components being either autoregressive or moving-average, see the Supplement for more details and an assessment of the similarity of the supLM statistics \citep{Gor21SM}. Finally, note that the tabulated values match those of table~1 of \cite{And03} where $\pi_0=0.25$.

\begin{table}
\caption{Tabulated quantiles for the asymptotic null distribution of the supLM statistics for the threshold range 25th-75th percentiles. The first two columns denote the AR and the MA orders, respectively.}\label{tab:q1}
\centering

\begin{tabular}{ccrrrrrrrr}
&& \multicolumn{4}{c}{$\sLM$} & \multicolumn{4}{c}{$\sLMg$} \\
      \cmidrule(lr){3-6}  \cmidrule(lr){7-10}
  AR & MA & 90\% & 95\% & 99\% & 99.9\% & 90\% & 95\% & 99\% & 99.9\% \\
      \cmidrule(lr){3-6}  \cmidrule(lr){7-10}
  1 & 1 &  9.61 & 11.37 & 15.19 & 20.38 & 11.64 & 13.44 & 17.42 & 22.83\\
  2 & 1 & 11.53 & 13.41 & 17.22 & 22.17 & 13.48 & 15.46 & 19.63 & 25.60\\
  3 & 1 & 13.74 & 15.71 & 19.98 & 25.04 & 15.59 & 17.61 & 21.91 & 27.98\\
  4 & 1 & 15.65 & 17.68 & 22.25 & 27.44 & 17.42 & 19.52 & 24.02 & 29.94\\
      \cmidrule(lr){3-6} \cmidrule(lr){7-10}
  1 & 2 &  9.64 & 11.47 & 15.50 & 20.25 & 13.69 & 15.57 & 19.67 & 25.07\\
  2 & 2 & 11.71 & 13.48 & 17.61 & 22.49 & 15.59 & 17.58 & 21.95 & 28.17\\
  3 & 2 & 13.46 & 15.35 & 19.33 & 25.06 & 17.13 & 19.18 & 23.54 & 29.78\\
  4 & 2 & 15.55 & 17.58 & 21.82 & 27.80 & 18.97 & 21.21 & 26.42 & 31.92\\
      \cmidrule(lr){3-6} \cmidrule(lr){7-10}
\end{tabular}
\end{table}
%
%
\section{The distribution under local alternatives and consistency of the test}\label{sec:alternatives}

In this section, we derive the asymptotic distribution of $T_n$ under a sequence of local alternatives and prove the consistency of the associated tests. For each $n$, the null hypothesis $H_{0,n}$ states that $\{X_t, t=0,\dots,n\}$  follows the model:
$$
X_t=\phi_{10}+\sum_{k=1}^{p}\phi_{1k}X_{t-k}-\sum_{s=1}^{q}\theta_{1s}\eps_{t-s}+\eps_t.
$$
The alternative hypothesis $H_{1,n}$ states that $\{X_t, t=0,\dots,n\}$  follows the model:
\begin{align}\label{eqn:mod_loc_alt}
  &X_t=\phi_{10}+\sum_{k=1}^{p}\phi_{1k}X_{t-k}-\sum_{s=1}^{q}\theta_{1s}\eps_{t-s} +\eps_t + \nonumber\\
  &\left[\left(\phi_{10}+\frac{h_{10}}{\sqrt{n}}\right)+\sum_{k=1}^{p} \left(\phi_{1k}+\frac{h_{1k}}{\sqrt{n}}\right)X_{t-k}-\sum_{s=1}^{q} \left(\theta_{1s}+\frac{h_{2s}}{\sqrt{n}}\right)\eps_{t-s}\right]I_{r_0}(X_{t-d}).
\end{align}
\noindent
where $\mathbf{h}=\left(h_{10},h_{11},\cdots,h_{1p},h_{21},\cdots,h_{2q}\right)^\intercal\in\mathds{R}^{p+q+1}$ is a fixed vector and $r_0$ is a fixed scalar. In the case of the $\sLM$ statistic, the rightmost summation term within square brackets is absent so that $\mathbf{h}=\left(h_{10},h_{11},\cdots,h_{1p}\right)^\intercal\in\mathds{R}^{p+1}$.
\par
Let $P_{0,n}$ and $P_{1,n}$ be the probability measure of $\left(X_0,X_1,\dots, X_n\right)$ under $H_{0,n}$ and $H_{1,n}$, respectively. In the following proposition we prove the asymptotic normality of the loglikelihood ratio and the contiguity of $P_{1,n}$ to $P_{0,n}$. As in C5 of Chan (1990), 3.1 of Ling and Tong (2005), A5 of Li and Li (2011), we assume

\begin{description}
  \item[A3.] The density $f$ of $\eps_t$ is absolutely continuous with derivative $f^\prime$ almost everywhere and
$
\int \left(\frac{f^\prime(x)}{f(x)}\right)^2 f(x)\,dx < \infty.
$
\end{description}

%
\begin{proposition}\label{prop:lp}
  Under assumptions A1-A3, it holds that:
  \begin{description}
    \item [(i)] Let $\nabla_2(r)$ be a Gaussian distributed random vector with zero mean and covariance matrix equal to $\Lambda_{22}(r)$. Under the null hypothesis, the log-likelihood ratio $\log\frac{d P_{1,n}}{d P_{0,n}}$ converges to the Gaussian random variable
    $$\mathbf{h}^\intercal\nabla_2(r_0)-\frac{1}{2}\mathbf{h}^\intercal\Lambda_{22}(r_0)\mathbf{h}.$$
    \item [(ii)] $\{P_{1,n}\}$ is contiguous to $\{P_{0,n}\}$.
  \end{description}
\end{proposition}
\noindent
Next, we derive the asymptotic distribution of $T_n$ under a sequence of local alternatives $H_{1,n}$:
\begin{theorem}\label{th:loc_distr}
  Assume A1-A3 to hold. Under $H_{1,n}$, it holds that:
  \begin{description}
    \item [(i)] $T_n$ converges weakly in $\mathcal{D}(-\infty,\infty)$ to $$(\xi_r+\gamma_r)^\intercal\left(\Lambda_{22}(r)-\Lambda_{2,1}(r)\Lambda^{-1}_{11}\Lambda_{12}(r)\right)^{-1}(\xi_r+\gamma_r),$$ where $\gamma_r=\left\{\Lambda_{22}(\min\{r,r_0\})-\Lambda_{21}(r)\Lambda^{-1}_{11}\Lambda_{12}(r_0)\right\}\mathbf h$.
    \item [(ii)] $\sup_{r\in[r_L,r_U]} T_n$ converges to
    $$\sup_{r\in[r_L,r_U]}(\xi_r+\gamma_r)^\intercal\left(\Lambda_{22}(r)-\Lambda_{21}(r)\Lambda^{-1}_{11}\Lambda_{12}(r)\right)^{-1}(\xi_r+\gamma_r).$$
  \end{description}
\end{theorem}
\noindent
Finally, we prove the consistency of our tests.
\begin{theorem}\label{th:power}
Under $H_{1,n}$, as $|\mathbf h|\to\infty$, the test statistic $T_n$  has power approaching 100\%.
\end{theorem}
Note that Proposition~\ref{prop:lp} and Theorem~\ref{th:loc_distr} can be proved for the $\sLM$ statistic without assumption A3. The two proofs are reported separately in the Supplementary Material.


\section{Finite-sample performance}\label{sec:finite_performance}

In this section, we investigate the finite sample performance of our supLM tests ($\sLM$, $\sLMg$) and compare them with the quasi-likelihood ratio test developed in \cite{Li11} (qLR). Hereafter $\eps_t$, $t=1,\dots,n$ is generated from a standard Gaussian white noise, the length of the series is $n=100,200,500$, the nominal size is $\alpha=0.05$ and the number of Monte Carlo replications is 1000. For our tests we have used the tabulated values of Table~\ref{tab:q1}. For the qLR test we have used $B=1000$ resamples. In Section~\ref{sec:size}, we study the size of the tests; Section~\ref{sec:power} shows the power of the tests in scenarios where $i)$ only the autoregressive parameters change across regimes, $ii)$ only the moving-average parameters change across regimes and $iii)$ both the autoregressive and the moving-average parameters change across regimes. Then, we assess the behaviour of the tests in presence of model misspecification (Section~\ref{sec:ms}) and when the order of the ARMA process tested is treated as unknown and is selected by means of the Hannan-Rissanen method (Section~\ref{sec:modsel}).

\subsection{Size of the tests}\label{sec:size}

\begin{table}
\caption{Empirical size at nominal level $5\%$ of the supLM tests ($\sLM$, $\sLMg$) and the quasi-Likelihood Ratio test (qLR). Rejection percentages from the ARMA$(1,1)$ model of Eq.~(\ref{eqn:ARMA_fsp}). Sample size $n=100,200,500$.}\label{tab:1}
\centering 
\begin{tabular}{rrrrrrrrrrr}
&& \multicolumn{3}{c}{$n=100$}& \multicolumn{3}{c}{$n=200$}& \multicolumn{3}{c}{$n=500$}\\
    \cmidrule(lr){3-5} \cmidrule(lr){6-8} \cmidrule(lr){9-11}
 $\phi$ & $\theta$  & $\sLM$ & $\sLMg$ & qLR & $\sLM$ & $\sLMg$ & qLR & $\sLM$ & $\sLMg$ & qLR \\
    \cmidrule(lr){3-5} \cmidrule(lr){6-8} \cmidrule(lr){9-11}
  -0.6&-0.8 & 5.8 &10.0 & 20.2 & 4.6 & 6.2 &  7.0 & 4.4 & 5.7 & 5.3 \\
  -0.3&-0.8 & 3.4 & 8.4 & 20.2 & 3.4 & 7.0 &  9.7 & 4.9 & 5.3 & 3.3 \\
   0.0&-0.8 & 4.3 & 7.4 & 20.4 & 4.1 & 7.2 &  8.7 & 4.2 & 4.2 & 5.2 \\
   0.3&-0.8 & 4.2 & 8.7 & 24.5 & 5.0 & 8.0 & 10.5 & 4.9 & 4.6 & 4.4 \\
   0.6&-0.8 & 4.6 & 8.4 & 32.0 & 5.6 & 7.9 & 22.2 & 4.4 & 5.8 & 7.6 \\
    \cmidrule(lr){3-5} \cmidrule(lr){6-8} \cmidrule(lr){9-11}
 -0.6&-0.4  & 6.1 & 8.5 &  5.9 & 5.1 & 5.2 & 3.7 & 4.7 & 4.8 & 3.1 \\
 -0.3&-0.4  & 7.5 & 8.3 &  9.7 & 4.9 & 5.0 & 5.1 & 5.9 & 5.8 & 3.8 \\
  0.0&-0.4  & 4.5 & 4.2 & 24.2 & 4.1 & 5.2 & 8.2 & 5.0 & 4.5 & 5.7 \\
  0.3&-0.4  & 5.4 & 6.2 & 45.6 & 3.8 & 4.3 &38.1 & 4.3 & 4.3 &23.4 \\
  0.6&-0.4  & 7.1 & 6.5 & 27.6 & 4.4 & 5.6 & 8.8 & 6.1 & 5.2 & 3.2 \\
    \cmidrule(lr){3-5} \cmidrule(lr){6-8} \cmidrule(lr){9-11}
   -0.6&0.0 & 4.6 & 4.2 & 10.0 & 4.2 & 4.0 & 5.8 & 4.6 & 4.9 & 3.1 \\
   -0.3&0.0 & 5.1 & 5.6 & 31.8 & 3.4 & 3.7 &16.0 & 5.9 & 6.2 & 4.8 \\
    0.0&0.0 & 9.4 &11.4 & 53.5 & 9.2 & 9.4 &42.0 & 6.7 & 7.1 &34.0 \\
    0.3&0.0 & 7.5 & 6.9 & 22.4 & 5.1 & 4.6 & 7.6 & 3.3 & 4.2 & 4.3 \\
    0.6&0.0 & 9.8 & 8.5 &  7.5 & 6.1 & 5.4 & 4.1 & 5.2 & 5.8 & 4.6 \\
    \cmidrule(lr){3-5} \cmidrule(lr){6-8} \cmidrule(lr){9-11}
   -0.6&0.4 & 4.2 & 4.8 & 41.0 & 3.3 & 4.3 &26.2 & 4.6 & 4.4 &13.9 \\
   -0.3&0.4 & 4.7 & 5.6 & 46.1 & 4.1 & 5.0 &40.7 & 3.9 & 4.5 &24.0 \\
    0.0&0.4 & 5.0 & 4.5 & 20.4 & 3.7 & 4.1 & 8.2 & 4.5 & 5.2 & 5.1 \\
    0.3&0.4 & 6.4 & 7.7 & 10.8 & 6.0 & 7.1 & 6.0 & 5.1 & 5.4 & 4.7 \\
    0.6&0.4 &15.5 &14.4 &  6.7 & 8.8 & 9.2 & 5.8 & 6.9 & 7.7 & 3.2 \\
    \cmidrule(lr){3-5} \cmidrule(lr){6-8} \cmidrule(lr){9-11}
   -0.6&0.8 & 10.3 & 16.8 & 45.0 & 7.2 & 10.7 & 25.2 & 4.5 & 7.0 &10.2 \\
   -0.3&0.8 &  8.3 & 14.8 & 29.1 & 6.3 & 10.4 & 14.9 & 4.2 & 6.6 & 7.3 \\
    0.0&0.8 &  8.2 & 16.0 & 22.4 & 6.0 & 11.3 &  9.2 & 4.9 & 7.2 & 6.3 \\
    0.3&0.8 &  7.7 & 14.4 & 21.0 & 7.2 & 10.0 &  9.8 & 4.1 & 6.5 & 5.6 \\
    0.6&0.8 & 11.9 & 16.0 & 19.5 & 9.3 & 11.2 &  8.1 & 6.3 & 8.1 & 4.9 \\
    \cmidrule(lr){3-5} \cmidrule(lr){6-8} \cmidrule(lr){9-11}
\end{tabular}
\end{table}
We have generated time series from 25 different simulation settings of the following ARMA$(1,1)$ model:
\begin{equation}\label{eqn:ARMA_fsp}
  X_t=\phi_{11} X_{t-1}+\eps_t-\theta_{11}\eps_{t-1}
\end{equation}
where $\phi_{11}=0,\pm 0.3, \pm 0.6$ and $\theta_{11}=0,\pm 0.4, \pm 0.8$. Table~\ref{tab:1} shows the rejection percentages for the three sample sizes in use. Note that the case $\theta_{11} = 0$ corresponds to testing an AR versus a TAR model. For $n=100$ the qLR test is biased in almost all settings and reaches 53\% of false rejections for the case $\theta_{11}=\phi_{11}=0$. The size of the $\sLM$ test is always acceptable as it is slightly greater than 10\% only in three cases and its maximum value is 15.5\%. The size of the $\sLMg$ is slightly more biased than that of the $\sLM$ test. When $n=200$ the bias of the $\sLM$ test reduces further and its size is not far from the nominal 5\% in most situations. This also holds for the $\sLMg$ test, except for the case $\theta_{11}=0.8$ where the size is still around 10\%. This is not the case for the qLR test whose size is close to 40\% in three simulation settings. When $n=500$ both our supLM tests achieve a size which is close to the nominal 5\% level, whereas the qLR test is still severely biased for some cases when near cancellation occurs, particularly when $\theta_{11}=\phi_{11}=0$ . One may argue that it is not appropriate to apply these tests to a realization of a white noise process and it would be more sensible to apply other kinds of tests in first place. Nevertheless, it may occur that a threshold process is mistaken for a white noise if the piecewise linear structure is such that the parameters of a linear ARMA fit result non-significant. Indeed, some sort of mis-specification is always present and this aspect will be investigated in Section~\ref{sec:ms}.
%

\subsection{Power of the tests}\label{sec:power}
%
In this section we study the power of the supLM tests and highlight the differences between them. Note that the parameter vector $\bPsi$ (see Eq.\ref{eq:pars}) represents the departure from the null hypothesis and in all the simulations below we take sequences of increasing distance from $H_0$ in all of its components. We simulate from three different TARMA$(1,1)$ models where $i)$ only the autoregressive parameters change across regimes, $ii)$ only the moving-average parameters change across regimes, and $iii)$ both the autoregressive and the moving-average parameters change across regimes. As for the first case, we simulate from the following model:
\begin{equation}\label{TARMAsim}
 X_t = -0.5 - 0.2 X_{t-1} - \theta_{11} \eps_{t-1} + \left(\Psi_{10} + \Psi_{11} X_{t-1}\right)I(X_{t-1}\leq 0) + \eps_t.
\end{equation}
where $\Psi_{10}$,$\Psi_{11}$ are as in Table~\ref{tab:p1}, first two columns. We combine these with $\theta_{11}=0,\pm 0.4, \pm 0.8$ as to obtain 20 different parameter settings. Table~\ref{tab:p1} presents the size-corrected power of the tests (in percentage). Clearly, the supLM tests outperform the qLR test uniformly (except for a single case). Note that the power depends upon the true value of $\theta_{11}$ and the case $\theta_{11}=0$ seems to impinge most negatively. In such instance, the qLR test has no power even for $n=200$ whereas the supLM tests show power loss due to the size correction only for $n=100$. Overall, starting from $n=200$ both the supLM tests present a good power in almost every situation. As expected, the $\sLM$ test is slightly superior to the $\sLMg$ test since the moving-average parameter is fixed across regimes.

\begin{table}
\centering
\caption{Size-corrected power for the TARMA$(1,1)$ model of Eq.~(\ref{TARMAsim}), case $i)$: only the autoregressive parameters change across regimes. Sample size $n=100,200,500$, $\alpha=5\%$.}\label{tab:p1}
\begin{tabular}{rrrrrrrrrrrr}
&&& \multicolumn{3}{c}{$n=100$}& \multicolumn{3}{c}{$n=200$}& \multicolumn{3}{c}{$n=500$}\\
    \cmidrule(lr){4-6} \cmidrule(lr){7-9} \cmidrule(lr){10-12}
$\Psi_{10}$ & $\Psi_{11}$ & $\theta_{11}$ & $\sLM$ & $\sLMg$ & qLR & $\sLM$  & $\sLMg$ & qLR    & $\sLM$ & $\sLMg$ & qLR \\
    \cmidrule(lr){4-6} \cmidrule(lr){7-9} \cmidrule(lr){10-12}
  0.1 & 0.4 & -0.8 & 27.1 & 22.7 & 12.3 & 57.4 & 47.2 & 44.7 &  96.6 &  95.7 &  95.4 \\
  0.3 & 0.6 & -0.8 & 61.2 & 51.3 & 28.9 & 93.9 & 91.1 & 87.7 & 100.0 & 100.0 & 100.0 \\
  0.5 & 0.8 & -0.8 & 89.3 & 83.7 & 55.0 & 99.8 & 99.3 & 98.5 & 100.0 & 100.0 & 100.0 \\
  0.7 & 1.0 & -0.8 & 96.6 & 94.6 & 74.0 &100.0 & 99.9 & 99.7 & 100.0 & 100.0 & 100.0 \\
    \cmidrule(lr){4-6} \cmidrule(lr){7-9} \cmidrule(lr){10-12}
  0.1 & 0.4 & -0.4 & 12.6 &  9.2 &  0.0 & 22.1 & 22.2 &  2.3 &  65.1 &  64.5 &  45.6 \\
  0.3 & 0.6 & -0.4 & 30.8 & 24.5 &  0.0 & 63.8 & 63.9 & 14.7 &  99.0 &  98.9 &  96.3 \\
  0.5 & 0.8 & -0.4 & 63.0 & 54.6 &  0.0 & 95.6 & 96.2 & 53.7 & 100.0 & 100.0 & 100.0 \\
  0.7 & 1.0 & -0.4 & 88.8 & 86.3 &  0.0 & 99.3 & 99.5 & 86.9 & 100.0 & 100.0 & 100.0 \\
    \cmidrule(lr){4-6} \cmidrule(lr){7-9} \cmidrule(lr){10-12}
  0.1 & 0.4 &  0.0 & 11.9 & 10.2 &  0.0 & 24.1 & 19.9 &  0.0 &  42.5 &  40.1 &  37.1 \\
  0.3 & 0.6 &  0.0 & 26.9 & 23.5 &  0.0 & 57.4 & 50.4 &  0.0 &  92.4 &  90.7 &  73.3 \\
  0.5 & 0.8 &  0.0 & 46.0 & 41.4 &  0.0 & 91.2 & 87.9 &  0.0 & 100.0 & 100.0 &  98.7 \\
  0.7 & 1.0 &  0.0 & 75.8 & 72.0 &  0.0 & 99.8 & 99.0 &  0.0 & 100.0 & 100.0 & 100.0 \\
    \cmidrule(lr){4-6} \cmidrule(lr){7-9} \cmidrule(lr){10-12}
  0.1 & 0.4 &  0.4 & 14.0 & 13.1 & 24.1 & 31.5 & 30.9 & 25.3 &  66.6 &  61.7 &  24.0 \\
  0.3 & 0.6 &  0.4 & 33.1 & 29.9 & 41.8 & 70.6 & 70.4 & 55.7 &  98.9 &  98.5 &  84.7 \\
  0.5 & 0.8 &  0.4 & 57.5 & 52.9 & 58.7 & 91.8 & 91.5 & 81.0 &  99.4 &  99.4 &  99.2 \\
  0.7 & 1.0 &  0.4 & 70.0 & 66.4 & 70.0 & 93.3 & 93.8 & 93.4 &  98.5 &  98.1 &  99.2 \\
    \cmidrule(lr){4-6} \cmidrule(lr){7-9} \cmidrule(lr){10-12}
  0.1 & 0.4 &  0.8 & 33.8 & 21.5 &  4.8 & 76.4 & 71.2 & 10.7 &  99.9 &  99.7 &  47.3 \\
  0.3 & 0.6 &  0.8 & 72.1 & 51.2 & 11.4 & 99.5 & 98.9 & 33.2 & 100.0 & 100.0 &  94.2 \\
  0.5 & 0.8 &  0.8 & 92.2 & 77.7 & 25.4 &100.0 &100.0 & 68.1 & 100.0 & 100.0 &  99.1 \\
  0.7 & 1.0 &  0.8 & 95.2 & 88.6 & 52.1 & 99.8 & 99.7 & 89.2 & 100.0 & 100.0 & 100.0 \\
    \cmidrule(lr){4-6} \cmidrule(lr){7-9} \cmidrule(lr){10-12}
\end{tabular}
\end{table}
\par
The case $ii)$ where only the moving-average parameter changes across regimes is studied by simulating from the following model:
\begin{equation}\label{TARMAsim2}
 X_t = \phi_{10} + \phi_{11} X_{t-1} - \theta_{11} \eps_{t-1} + \left(\Psi_{21}\eps_{t-1}\right)I(X_{t-1}\leq 0) + \eps_t.
\end{equation}
where $\theta_{11}=-0.6$, whereas $\phi_{10}$, $\phi_{11}$ and $\Psi_{21}$ are as in Table~\ref{tab:p2}, that shows the rejection percentages. In this case the behaviour of the tests depends on different factors. When the departure from the null hypothesis is mild, the qLR test has an advantage over supLM tests. The situation is reversed when $\Psi_{21}$ is large (e.g. $\Psi_{21}=1.2$): in such case both supLM tests are more powerful than the qLR test. When $n = 500$ and $\phi_0=0.6$, $\phi_1=0.7$ the $\sLMg$ test is always more powerful than the qLR test, whereas in the remaining cases there is not a clear winner and the results are comparable.
%
\begin{table}
\caption{Size-corrected power for the TARMA$(1,1)$ model of Eq.~(\ref{TARMAsim2}), case $ii)$: only the moving-average parameters change across regimes. Sample size $n=100,200,500$, $\alpha=5\%$.}\label{tab:p2}
\centering
\begin{tabular}{rrrrrrrrrrrr}
&&& \multicolumn{3}{c}{$n=100$}& \multicolumn{3}{c}{$n=200$}& \multicolumn{3}{c}{$n=500$}\\
    \cmidrule(lr){4-6} \cmidrule(lr){7-9} \cmidrule(lr){10-12}
$\phi_{10}$ & $\phi_{11}$ & $\Psi_{21}$  & $\sLM$ & $\sLMg$ & qLR & $\sLM$  & $\sLMg$ & qLR    & $\sLM$ & $\sLMg$ & qLR \\
    \cmidrule(lr){4-6} \cmidrule(lr){7-9} \cmidrule(lr){10-12}
  0.2 & 0.1 & 0.4 &  9.1 & 12.6 & 14.9 & 23.1 & 29.1 & 43.8 &  62.6 &  77.4 & 83.1 \\
  0.2 & 0.1 & 0.6 & 23.8 & 24.8 & 29.9 & 50.2 & 55.6 & 69.6 &  95.1 &  97.8 & 99.2 \\
  0.2 & 0.1 & 0.8 & 39.2 & 38.9 & 46.8 & 81.8 & 81.8 & 90.9 & 100.0 & 100.0 &100.0 \\
  0.2 & 0.1 & 1.2 & 71.2 & 67.7 & 75.2 & 96.4 & 96.6 & 94.5 &  99.9 &  99.9 & 99.7 \\
    \cmidrule(lr){4-6} \cmidrule(lr){7-9} \cmidrule(lr){10-12}
  0.4 & 0.4 & 0.4 &  8.3 & 10.3 & 15.5 & 23.3 & 34.2 & 47.0 &  62.6 &  85.1 & 91.4 \\
  0.4 & 0.4 & 0.6 & 17.2 & 20.4 & 31.4 & 52.3 & 62.9 & 73.6 &  95.4 &  98.8 & 99.6 \\
  0.4 & 0.4 & 0.8 & 38.3 & 39.1 & 47.6 & 83.8 & 87.3 & 91.6 & 100.0 & 100.0 & 99.9 \\
  0.4 & 0.4 & 1.2 & 77.5 & 72.4 & 75.5 & 99.4 & 99.3 & 98.7 & 100.0 & 100.0 &100.0 \\
    \cmidrule(lr){4-6} \cmidrule(lr){7-9} \cmidrule(lr){10-12}
  0.6 & 0.7 & 0.4 &  8.1 &  8.2 & 10.5 & 11.1 & 14.9 & 21.6 &  42.6 &  57.2 & 56.8 \\
  0.6 & 0.7 & 0.6 & 12.2 & 14.8 & 17.8 & 31.2 & 36.4 & 38.8 &  82.0 &  90.5 & 85.3 \\
  0.6 & 0.7 & 0.8 & 23.2 & 25.5 & 27.2 & 58.3 & 61.8 & 55.9 &  98.5 &  98.8 & 93.4 \\
  0.6 & 0.7 & 1.2 & 48.6 & 48.4 & 38.4 & 90.6 & 89.5 & 76.5 & 100.0 & 100.0 & 99.1 \\
    \cmidrule(lr){4-6} \cmidrule(lr){7-9} \cmidrule(lr){10-12}
\end{tabular}
\end{table}
\par
As concerns case $iii)$ we simulate from the following model:
\begin{equation}\label{TARMAsim3}
 X_t = \phi_{10} + \phi_{11} X_{t-1} - \theta_{11} \eps_{t-1} + \left(\Psi_{10} + \Psi_{11} X_{t-1} + \Psi_{21} \eps_{t-1}\right)I(X_{t-1}\leq 0) + \eps_t.
\end{equation}
where $\phi_{10}=-0.5$, $\phi_{11}=-0.5$, $\theta_{11}=0.5$ and $\Psi_{10}$, $\Psi_{11}$ and $\Psi_{21}$ are as in Table~\ref{tab:p3}, first three columns. When $n=100$ the $\sLMg$ and the qLR tests are comparable and more powerful than the $\sLM$ test. However when $n=200$ the $\sLMg$ test is more powerful than the qLR and $\sLM$ tests, whose power is comparable. When $n=500$ the $\sLMg$ test is always the most powerful of the three. Also, on average the $\sLM$ test has 6\% less power than the qLR test. Note that, in principle, this setting is more favourable to the $\sLMg$ and qLR tests since, even if the sequence of departures from $H_0$ is monotonically increasing with respect to all the components, the rate is faster along the moving-average component $\Psi_{21}$ and slower on the autoregressive part $\Psi_{10}$ and $\Psi_{11}$ (see the first three columns of Table~\ref{tab:p3}). When all the components are distant from the null hypothesis, then, the power of the tests are either comparable or the $\sLM$ test is even more powerful (results not shown here).

\begin{table}
\caption{Size-corrected power for the TARMA$(1,1)$ model of Eq.~(\ref{TARMAsim3}), case $iii)$: both the autoregressive and the moving-average parameters change across regimes. Sample size $n=100,200,500$, $\alpha=5\%$.}\label{tab:p3}
\centering
\begin{tabular}{rrrrrrrrrrrr}
&&& \multicolumn{3}{c}{$n=100$}& \multicolumn{3}{c}{$n=200$}& \multicolumn{3}{c}{$n=500$}\\
    \cmidrule(lr){4-6} \cmidrule(lr){7-9} \cmidrule(lr){10-12}
$\Psi_{10}$ & $\Psi_{11}$ & $\Psi_{21}$ & $\sLM$ & $\sLMg$ & qLR & $\sLM$  & $\sLMg$ & qLR    & $\sLM$ & $\sLMg$ & qLR \\
    \cmidrule(lr){4-6} \cmidrule(lr){7-9} \cmidrule(lr){10-12}
  0.02 & 0.02 & -0.10 &  6.2 &  6.0 &  6.5 &  8.8 &  7.9 &  5.0 &  7.8 & 10.9 &  8.9 \\
  0.04 & 0.04 & -0.20 &  7.0 &  7.1 &  9.5 & 10.1 & 13.0 &  8.8 & 18.5 & 32.6 & 21.6 \\
  0.06 & 0.06 & -0.30 &  8.8 & 11.3 & 11.1 & 17.9 & 23.5 & 15.6 & 37.3 & 66.8 & 50.7 \\
  0.08 & 0.08 & -0.40 & 14.2 & 16.8 & 17.0 & 28.9 & 40.9 & 29.9 & 63.5 & 90.9 & 78.9 \\
  0.10 & 0.10 & -0.50 & 18.3 & 23.9 & 23.8 & 43.1 & 57.6 & 44.5 & 85.0 & 97.1 & 92.3 \\
  0.12 & 0.12 & -0.60 & 25.6 & 32.6 & 30.1 & 54.9 & 72.3 & 60.3 & 95.7 & 99.6 & 99.3 \\
  0.14 & 0.14 & -0.70 & 30.9 & 37.3 & 40.4 & 76.0 & 85.3 & 74.4 & 99.4 &100.0 & 99.8 \\
    \cmidrule(lr){4-6} \cmidrule(lr){7-9} \cmidrule(lr){10-12}
\end{tabular}
\end{table}
The results for higher order TARMA models confirm the above conclusions and some of these are reported in the Supplement \cite{Gor21SM}.
\subsection{Size and power in presence of mis-specification}\label{sec:ms}
%
In this section we assess the impact of model mis-specification upon the performance of the tests. The sources of mis-specification can be diverse: as above, we focus on testing the ARMA$(1,1)$ versus the TARMA$(1,1)$ specification but the data generating process is not encompassed within the two models. Loosely speaking, we are investigating the capability of the test to detect general departures from linearity beyond the direct comparison of two specific models. Ideally, if the data generating process falls within the class of linear processes we would want the test not to reject the null hypothesis. Likewise, if the data generating process is non-linear in some of its components, then we expect the test to reject the null hypothesis. In Table~\ref{tab:dgp} we show the list of linear and non-linear data generating processes used. The seven linear processes are not ARMA$(1,1)$ since they contain higher-order autoregressive or moving-average terms. In the second part of the table we show non-linear processes that cannot be encompassed within the two-regime TARMA$(1,1)$ specification. In particular, we simulate from TAR models with both higher autoregressive order and more than two regimes. Lastly, we generate from six non-linear models that do not belong to the TARMA class, such as non-linear moving-average (NLMA), bilinear (BIL),  exponential autoregressive (EXPAR) and deterministic chaos (NLAR).
\begin{table}
  \centering 
  \caption{Data generating processes used to investigate the size and power of the tests under model mis-specification. Unless otherwise stated $\{\eps_t\}$ follows a standard Gaussian white noise.}\label{tab:dgp}
\begin{tabular}{lll}
\cmidrule(lr){2-3}
\multirow{7}{*}{\rotatebox[origin=c]{90}{\textsc{linear}}}
&01. AR5 &
 $     X_{t}=-0.6 X_{t-1}-0.4 X_{t-2}-0.3 X_{t-3}-0.4  X_{t-4} - 0.5 X_{t-5} + \eps_t$
\\
&02. AR2.1 &
 $    X_t= 0.75 X_{t-1} - 0.125 X_{t-2} + \eps_t$
 \\
&03. AR2.2 &
 $    X_t=  1.35 X_{t-1} - 0.55 X_{t-2} + \eps_t$
 \\
&04. ARMA21.1 &
 $    X_t= 0.75 X_{t-1} - 0.125 X_{t-2} - 0.7 \eps_{t-1} + \eps_t$
 \\
&05. ARMA21.2 &
 $    X_t= 0.75 X_{t-1} - 0.125 X_{t-2} + 0.7 \eps_{t-1} + \eps_t$
 \\
&06. ARMA22 &
 $    X_t= 0.75 X_{t-1} - 0.125 X_{t-2} + 0.7 \eps_{t-1} - 0.4 \eps_{t-2} + \eps_t$
 \\
&07. MA2 &
 $    X_t=  0.7 \eps_{t-1} - 0.125 \eps_{t-2} + \eps_t$
\\
\cmidrule(lr){2-3}
&&\\
\multirow{9}{*}{\rotatebox[origin=c]{90}{\textsc{non-linear}}}
&08. TAR3 &
$ X_t=
  \begin{cases}
    0.3X_{t-1}-0.7X_{t-2}+0.6X_{t-3}+\eps_{t}, & \text{if }  X_{t-1}\leq 0 \\
    -0.3X_{t-1}+0.7X_{t-2}-0.6X_{t-3}+\eps_{t}, & \text{if } X_{t-1}>0
  \end{cases}
$
\\
&09. 3TAR1 &
$ X_t=
  \begin{cases}
    0.3+0.5 X_{t-1}+\eps_{t}, & \text{if } X_{t-1}\leq -1 \\
    0.3+X_{t-1}+\eps_{t},     & \text{if } -1 < X_{t-1}\leq 1\\
    0.3+0.5X_{t-1}+\eps_{t},  & \text{if } X_{t-1}> 1
  \end{cases}
$
\\
&10. NLMA.1 &
 $     X_t=  - 0.8\eps^2_{t-1} + \eps_{t}$
 \\
&11. NLMA.2 &
 $     X_t=  \phantom{-}0.8\eps^2_{t-1} + \eps_{t}$
\\
&12. BIL.1 &
 $     X_t=  0.5 - 0.4 X_{t-1} + 0.4 \eps_{t-1}X_{t-1} + \eps_{t}$
 \\
&13. BIL.2 &
 $     X_t=   0.7 \eps_{t-1}X_{t-2} + \eps_{t}$
 \\
&14. EXPAR.1 &
    $X_t = 0.3 + 10 \,\exp(-X^2_{t-1})X_{t-1} + \eps_t$
 \\
&15. EXPAR.2 &
    $X_t = 0.3 + 100 \,\exp(-X^2_{t-1})X_{t-1} + \eps_t$
\\
&16. NLAR &
    $X_t = 4\, X_t \, (1-X_t)$
\\
\cmidrule(lr){2-3}
\end{tabular}
\end{table}
\begin{table}
\caption{Rejection percentages under model misspecification for the processes of Table~\ref{tab:dgp} for the supLM tests ($\sLM$, $\sLMg$) and the quasi-Likelihood Ratio test (qLR). The upper panel (linear processes) reflects the empirical size at nominal level $5\%$. The lower panel (non-linear processes) reflects the empirical power for non-linear processes that are not representable as a TARMA$(1,1)$ process.}\label{tab:3}
\centering 
\begin{tabular}{rrrrrrrrrrr}
&& \multicolumn{3}{c}{$n=100$}& \multicolumn{3}{c}{$n=200$}& \multicolumn{3}{c}{$n=500$}\\
   \cmidrule(lr){3-5} \cmidrule(lr){6-8} \cmidrule(lr){9-11}
&& $\sLM$ & $\sLMg$ & qLR & $\sLM$ & $\sLMg$ & qLR & $\sLM$ & $\sLMg$ & qLR \\
   \cmidrule(lr){3-5} \cmidrule(lr){6-8} \cmidrule(lr){9-11}
\multirow{7}{*}{\rotatebox[origin=c]{90}{\textsc{linear}}}&
   AR5       &  7.9 & 15.2 & 22.7 &  5.5  &  9.8 &  7.8 &  4.6 &  6.9 &  4.6 \\
&  AR2.1     &  6.7 &  7.1 &  6.8 &  4.7  &  5.5 &  4.0 &  5.3 &  5.5 &  4.0 \\
&  AR2.2     &  8.7 & 10.1 &  6.6 &  4.1  &  5.8 &  3.2 &  3.4 &  5.0 &  1.6 \\
&  ARMA21.1  &  5.9 &  6.8 & 48.0 &  5.5  &  5.2 & 39.7 &  5.5 &  4.8 & 27.8 \\
&  ARMA21.2  &  5.8 &  8.9 & 15.8 &  4.5  &  5.7 &  8.3 &  4.0 &  4.9 &  4.1 \\
&  ARMA22    &  5.5 & 13.6 & 29.1 &  3.7  & 12.7 & 17.4 &  3.5 &  7.5 &  8.5 \\
&  MA2       &  3.5 & 12.1 & 29.9 &  2.3  &  5.8 & 17.0 &  4.4 &  6.2 &  6.5 \\
   \cmidrule(lr){3-5} \cmidrule(lr){6-8} \cmidrule(lr){9-11}
\multirow{10}{*}{\rotatebox[origin=c]{90}{\textsc{non-linear}}}&
     TAR3    & 34.7 & 98.1 & 63.0 &  61.8 & 99.7 & 48.0 & 96.7 &100.0 & 36.8 \\
&    3TAR1   & 19.2 & 18.4 &  8.8 &  36.1 & 30.4 &  8.5 & 78.5 & 72.4 & 16.5 \\
&    NLMA.1  & 85.1 & 84.0 & 78.7 &  97.3 & 97.2 & 92.1 & 99.8 & 99.9 & 99.0 \\
&    NLMA.2  & 86.4 & 86.8 & 73.6 &  98.3 & 98.5 & 85.2 &100.0 & 99.9 & 97.0 \\
&    BIL.1   & 12.0 & 62.1 & 53.2 &  13.5 & 84.8 & 58.2 & 14.7 & 95.0 & 77.4 \\
&    BIL.2   & 84.0 & 83.1 & 86.8 &  98.7 & 98.9 & 96.0 &100.0 &100.0 & 99.9 \\
&    EXPAR.1 &100.0 &100.0 & 32.5 & 100.0 &100.0 & 30.5 &100.0 &100.0 & 80.1 \\
&    EXPAR.2 & 95.8 & 99.1 & 45.0 &  99.6 & 99.9 & 59.3 &100.0 &100.0 & 96.2 \\
&    NLAR    &100.0 &100.0 & 63.7 & 100.0 &100.0 & 47.0 &100.0 &100.0 & 27.1 \\
   \cmidrule(lr){3-5} \cmidrule(lr){6-8} \cmidrule(lr){9-11}
\end{tabular}
\end{table}
The rejection percentages are reported in Table~\ref{tab:3}. As discussed above, the first seven rows should reflect the empirical size at nominal level $5\%$ under mis-specification. Consistently with the results of Table~\ref{tab:1}, the $\sLM$ test is well behaved in terms of size even for $n=100$ whereas the $\sLMg$ test has acceptable size starting from $n=200$. The qLR test presents acceptable size just for $n=500$, except for the ARMA21.1 case with a 27.8\% of false rejections. The lower panel of Table~\ref{tab:3} shows the rejection percentages for the 9 non-linear processes that do not belong to the TARMA$(1,1)$ class. Here, the two supLM tests show higher power in almost every situation even for $n=100$ and with a consistent increase over the sample size. The qLR test has good power in several instances but for the TAR3, the 3TAR1 and the NLAR processes the power decreases as the sample size increases. The causes of this phenomenon are not clear and deserve further investigation, we offer some discussion in the Conclusions section. The general conclusions that can be drawn are that the supLM tests are robust with respect to modelling mis-specifications, both in terms of size and power. The same cannot be said for the qLR test that, in some cases presents oversize and power loss even for $n=500$.
%
\subsection{The impact of model selection}\label{sec:modsel}
%
\begin{table}
\caption{Empirical size of the supLM tests at nominal level 5\% for 6 parameterizations of the ARMA$(2,2)$ process. The subscript ``\textsc{HR}'' indicates that the order of the ARMA model has been selected through the Hannan-Rissanen procedure.}\label{tab:ms1}
\centering

\begin{tabular}{rrrrrrrrrr}
&&&& \multicolumn{2}{c}{$n=100$}& \multicolumn{2}{c}{$n=200$}& \multicolumn{2}{c}{$n=500$}\\
    \cmidrule(lr){5-6} \cmidrule(lr){7-8} \cmidrule(lr){9-10}
$\phi_1$ & $\phi_2$ & $\theta_{1}$ & $\theta_{2}$ & $\sLM$ & $\sLMh$ & $\sLM$ & $\sLMh$ & $\sLM$ & $\sLMh$ \\
    \cmidrule(lr){5-6} \cmidrule(lr){7-8} \cmidrule(lr){9-10}
  -0.35 & -0.45 &  0.25 & -0.25 & 5.4 & 4.4 & 3.6 & 3.7 & 4.9 & 4.9 \\
   0.45 & -0.55 &  0.25 & -0.25 & 5.6 & 5.8 & 5.7 & 4.8 & 4.6 & 4.6 \\
  -0.90 & -0.25 &  0.25 & -0.25 & 6.1 & 4.6 & 5.8 & 5.9 & 4.4 & 5.6 \\
  -0.35 & -0.45 & -0.25 &  0.25 & 2.9 & 3.0 & 4.3 & 4.4 & 5.3 & 5.3 \\
   0.45 & -0.55 & -0.25 &  0.25 & 3.0 & 3.4 & 3.9 & 4.3 & 5.3 & 5.0 \\
  -0.90 & -0.25 & -0.25 &  0.25 & 7.1 & 4.7 & 6.8 & 4.6 & 5.7 & 5.2 \\
    \cmidrule(lr){5-6} \cmidrule(lr){7-8} \cmidrule(lr){9-10}\\
&&&& \multicolumn{2}{c}{$n=100$}& \multicolumn{2}{c}{$n=200$}& \multicolumn{2}{c}{$n=500$}\\
    \cmidrule(lr){5-6} \cmidrule(lr){7-8} \cmidrule(lr){9-10}
$\phi_1$ & $\phi_2$ & $\theta_{1}$ & $\theta_{2}$ & $\sLMg$ & $\sLMgh$ & $\sLMg$ & $\sLMgh$ & $\sLMg$ & $\sLMgh$ \\
    \cmidrule(lr){5-6} \cmidrule(lr){7-8} \cmidrule(lr){9-10}
  -0.35 & -0.45 &  0.25 & -0.25 & 6.1 & 4.3 & 4.3 & 3.7 & 4.5 & 4.9 \\
   0.45 & -0.55 &  0.25 & -0.25 & 7.2 & 6.4 & 4.8 & 5.3 & 4.5 & 5.1 \\
  -0.90 & -0.25 &  0.25 & -0.25 & 6.7 & 5.9 & 4.9 & 5.3 & 4.8 & 5.4 \\
  -0.35 & -0.45 & -0.25 &  0.25 & 3.1 & 3.0 & 4.9 & 4.8 & 4.6 & 4.2 \\
   0.45 & -0.55 & -0.25 &  0.25 & 4.4 & 4.8 & 4.1 & 4.7 & 4.9 & 4.8 \\
  -0.90 & -0.25 & -0.25 &  0.25 &11.4 & 6.5 & 6.7 & 4.3 & 5.3 & 5.2 \\
    \cmidrule(lr){5-6} \cmidrule(lr){7-8} \cmidrule(lr){9-10}
\end{tabular}
\end{table}
Testing the ARMA against the TARMA specification requires selecting a specific order beforehand. In this section we show that there is virtually no loss incurred in using our supLM tests when no previous information on the order is available, provided a proper model selection procedure is adopted. We advocate the use of the consistent ARMA order selection proposed in \cite{Han82} (see also \cite{Cho92}).
\par
In Table~\ref{tab:ms1} we present the empirical size of the supLM tests at nominal level 5\% for 6 parameterizations of an ARMA$(2,2)$ process (see the first four columns). The upper panel of the table refers to the $\sLM$ test whereas the lower panel refers to the $\sLMg$ test. In both cases, the subscript \textsc{HR} indicates that the order of the ARMA process has been selected through the Hannan-Rissanen procedure; the true order has been used otherwise. The results indicate that not only the model selection step does not produce a size bias, but it also seems to reduce it in some instances.
\par
The impact of model selection on the power of the tests is shown in Table~\ref{tab:ms2} where we simulate 12 parameter settings of the TARMA$(1,1)$ model of Eq.~(\ref{TARMAsim}). Clearly, the power loss produced by using model selection is minimal and lies within 2\% for the $\sLM$ test and 3\% for the $\sLMg$ test. Finally, also in presence of mis-specification the results of the HR model selection poses no problems. The results are shown in the Supplementary material.
%
%
\begin{table}
\caption{Empirical power for the TARMA$(1,1)$ model of Eq.~(\ref{TARMAsim}). Sample size $n=100,200,500$. The subscript ``\textsc{HR}'' indicates that the order of the ARMA model has been selected through the Hannan-Rissanen procedure.}\label{tab:ms2}
\centering
\begin{tabular}{rrrrrrrrr}
&&& \multicolumn{2}{c}{$n=100$}& \multicolumn{2}{c}{$n=200$}& \multicolumn{2}{c}{$n=500$}\\
    \cmidrule(lr){4-5} \cmidrule(lr){6-7} \cmidrule(lr){8-9}
$\Psi_{10}$ & $\Psi_{11}$ & $\theta_{11}$ & $\sLM$ & $\sLMh$ & $\sLM$ & $\sLMh$ & $\sLM$ & $\sLMh$ \\
    \cmidrule(lr){4-5} \cmidrule(lr){6-7} \cmidrule(lr){8-9}
  0.1 & 0.4 &-0.5 & 13.0 & 12.9 & 30.2 & 29.4 & 71.2 & 69.4 \\
  0.3 & 0.6 &-0.5 & 34.8 & 35.1 & 71.9 & 70.6 & 99.6 & 98.9 \\
  0.5 & 0.8 &-0.5 & 65.5 & 65.2 & 97.1 & 96.3 &100.0 &100.0 \\
  0.7 & 1.0 &-0.5 & 90.4 & 89.6 & 99.7 & 99.5 &100.0 &100.0 \\
    \cmidrule(lr){4-5} \cmidrule(lr){6-7} \cmidrule(lr){8-9}
  0.1 & 0.4 & 0.0 & 15.7 & 14.9 & 19.0 & 19.2 & 43.0 & 43.0 \\
  0.3 & 0.6 & 0.0 & 30.7 & 30.4 & 51.8 & 51.0 & 94.0 & 93.6 \\
  0.5 & 0.8 & 0.0 & 54.5 & 54.6 & 87.9 & 87.5 &100.0 & 99.9 \\
  0.7 & 1.0 & 0.0 & 81.3 & 80.5 & 99.1 & 99.1 &100.0 &100.0 \\
    \cmidrule(lr){4-5} \cmidrule(lr){6-7} \cmidrule(lr){8-9}
  0.1 & 0.4 & 0.5 & 17.4 & 16.0 & 35.9 & 33.8 & 77.9 & 75.8 \\
  0.3 & 0.6 & 0.5 & 41.3 & 38.7 & 78.2 & 73.1 & 99.9 & 99.6 \\
  0.5 & 0.8 & 0.5 & 70.7 & 67.2 & 95.7 & 93.9 & 99.9 & 99.7 \\
  0.7 & 1.0 & 0.5 & 75.5 & 74.6 & 93.7 & 93.1 & 99.5 & 99.5 \\
    \cmidrule(lr){4-5} \cmidrule(lr){6-7} \cmidrule(lr){8-9}\\
&&& \multicolumn{2}{c}{$n=100$}& \multicolumn{2}{c}{$n=200$}& \multicolumn{2}{c}{$n=500$}\\
    \cmidrule(lr){4-5} \cmidrule(lr){6-7} \cmidrule(lr){8-9}
$\Psi_{10}$ & $\Psi_{11}$ & $\theta_{11}$ & $\sLMg$ & $\sLMgh$ & $\sLMg$ & $\sLMgh$ & $\sLMg$ & $\sLMgh$ \\
    \cmidrule(lr){4-5} \cmidrule(lr){6-7} \cmidrule(lr){8-9}
  0.1 & 0.4 &-0.5 & 13.9 & 13.9 & 28.7 & 28.7 & 66.7 & 65.0 \\
  0.3 & 0.6 &-0.5 & 35.5 & 35.7 & 69.5 & 67.0 & 99.0 & 98.6 \\
  0.5 & 0.8 &-0.5 & 65.7 & 65.1 & 96.4 & 95.7 &100.0 &100.0 \\
  0.7 & 1.0 &-0.5 & 89.7 & 88.8 & 99.8 & 99.6 &100.0 &100.0 \\
    \cmidrule(lr){4-5} \cmidrule(lr){6-7} \cmidrule(lr){8-9}
  0.1 & 0.4 & 0.0 & 15.0 & 14.4 & 18.7 & 19.1 & 40.3 & 40.6 \\
  0.3 & 0.6 & 0.0 & 27.8 & 28.2 & 48.8 & 48.5 & 91.7 & 91.5 \\
  0.5 & 0.8 & 0.0 & 52.0 & 52.1 & 86.1 & 85.5 &100.0 & 99.9 \\
  0.7 & 1.0 & 0.0 & 80.1 & 79.7 & 98.9 & 98.7 &100.0 &100.0 \\
    \cmidrule(lr){4-5} \cmidrule(lr){6-7} \cmidrule(lr){8-9}
  0.1 & 0.4 & 0.5 & 16.4 & 16.0 & 32.6 & 31.2 & 75.4 & 73.6 \\
  0.3 & 0.6 & 0.5 & 40.8 & 38.7 & 75.3 & 71.6 & 99.6 & 98.9 \\
  0.5 & 0.8 & 0.5 & 68.3 & 65.8 & 95.9 & 94.5 &100.0 &100.0 \\
  0.7 & 1.0 & 0.5 & 74.8 & 74.3 & 93.7 & 93.0 & 99.6 & 99.6 \\
    \cmidrule(lr){4-5} \cmidrule(lr){6-7} \cmidrule(lr){8-9}
\end{tabular}
\end{table}%
\subsection{Discussion}
The Monte Carlo study has shown that supLM tests have good finite sample properties. They are also robust against model mis-specification and their performance is not affected if the order of the tested process is unknown, provided a consistent order selection procedure is used. The tests do not suffer from some of the drawbacks that affect the quasi-likelihood ratio test. The reasons can be diverse. First and foremost, supLM statistics only require fitting an ARMA model whereas the qLR test is bound to estimating a full TARMA model. We remind that there are no theoretical results regarding the sampling properties of the maximum likelihood estimators for the parameters of a TARMA model. Moreover, the qLR test by \cite{Li11} uses a representation in terms of a quadratic form but it is only valid asymptotically and this can impinge on the rate of convergence of the statistic towards its asymptotic distribution.
\par
The statistic $\sLM$, tests the ARMA($p,q$) against the TARMA($p,q$) model when only the $p$ autoregressive parameters change across regimes. However, the results show that such test has power also when only the moving-average parameters change. This could be ascribed to the duality between MA and AR processes and indicates a capability to detect general departures from linearity as also witnessed by the results of Section~\ref{sec:ms}. The results also show that, as expected, when either only the $q$ MA parameters or all the $p+q$ parameters of the ARMA model change across regimes, then the $\sLMg$ test is more powerful. The price to be paid for this superior power is the increased size bias in small samples. In general, we expect the two tests to behave similarly but in case of small samples the $\sLM$ statistic is recommended and can be used in conjunction with the test based upon $\sLMg$ .

\section{A real data application: the tree ring time series}\label{sec:real}
In this section we present an application of our test to the time series of the tree-ring standardized growth index. Tree rings provide a measure of the responses of tree growth to past climate variation and this information is very useful in climate studies. Despite the recognition that the climatic factors affecting tree growth form a complex network, according to the literature, the best model adopted is either the AR$(1)$ or the ARMA$(1,1)$, see table 3 in \citet{Fox01}. Usually, the indexes of many trees from the same site are used to cross date the rings and are finally averaged into a single index as to obtain a chronology that covers a long time span. Here we focus on the tree-ring chronology of a Pinus aristata var. longaeva (California, USA) from year 800 to 1979 ($n=1180$), for more details on the data see \cite{ca535}. 
\par
We test the ARMA$(1,1)$ specification against the following TARMA$(1,1)$ model
\begin{equation}\label{eq:fit.tarma}
  X_t = \begin{cases}
        \phi_{10}  + \phi_{11} X_{t-1} + \eps_t +  \theta_{11} \eps_{t-1}, & \text{ if } X_{t-1}\leq r \\
        \phi_{20}  + \phi_{21} X_{t-1} + \eps_t +  \theta_{11} \eps_{t-1}, & \text{ otherwise}.
      \end{cases}
\end{equation}
 With the threshold searched between the 10th to 90th percentiles, the $\sLM$ test statistic is 23.45, while the $\sLMg$ statistic is 25.21 and both of them correspond to a $p$-value smaller than 0.001, suggesting that tree-ring growth is regulated by floor. Table~\ref{tab:tr1} reports a TARMA$(1,1)$ model parameterized in the form of (\ref{eq:fit.tarma}) with common moving-average parameter but with unconstrained $\phi_{i,1}, i=0,1,2$, and an ARMA$(1,1)$ model fitted to the data. The estimated autoregressive parameters point to a threshold effect and the normalized AIC and BIC indicate an improvement with respect to the ARMA$(1,1)$ model. The estimated TARMA$(1,1)$ model is invertible and geometrically ergodic.
The estimated threshold is $\hat r = 0.97$ which is close to 1, the mean of the process, and identifies an upper regime where the growth is accelerated with respect to the lower regime. Finally, model diagnostics reported in the {Supplementary Material} indicate that the TARMA$(1,1)$ model provides a good fit to the data whereas an unaccounted dependence structure is present in the residuals of the ARMA$(1,1)$ model.
\begin{table}
  \centering
  \caption{Parameter estimates for the tree ring time series}\label{tab:tr1}
$
\begin{array}{cccccccccc}
              & \theta_{11}&\phi_{10}&\phi_{11}&\phi_{20}&\phi_{21}& r     & d   &\text{NAIC} & \text{NBIC} \\
  \midrule
  \text{ARMA} & -0.60      &  1.00    &  0.76    &          &          &       &     & 0.232  & 0.251 \\
              & (0.09)     & (0.01)   & (0.07)   &          &          &       &     &        &  \\
  \midrule
  \text{TARMA}& -0.44      &  0.54    &  0.37    &  0.29    &  0.71    & 0.97  & 1   & 0.213  & 0.240 \\
              & (0.09)     & (0.10)   & (0.11)   & (0.09)   & (0.09)   &       &     &        &  \\
\bottomrule
\end{array}
$
\end{table}
\section{Conclusions}\label{sec:conclusions}
In this paper we have presented consistent supremum Lagrange Multiplier tests to compare a linear ARMA specification against its TARMA extension. Our proposal extends previous results, such as \citet{Cha90a,Lin05} and enjoys very good finite-sample properties in terms of size and power. Moreover, being based upon asymptotic theory, it has a low computational burden. From the tabulated quantiles of the asymptotic distributions it seems that these depend only on the numbers of parameters tested and match those of \cite{And03} and this prompts interesting further theoretical investigations. The Monte Carlo study has shown that supLM tests are also robust against model mis-specification and their performance is not affected if the order of the tested process is unknown, provided a consistent order selection procedure is used. Our supLM tests do not suffer from some of the shortcomings that affect the quasi-likelihood ratio test so that they can be used for small samples. In such a case, the $\sLM$ statistic has less power than the $\sLMg$ statistic but it is better behaved in terms of size so that it is recommended. For sample sizes from 200 onwards, the two tests can be used in conjunction. The theoretical framework of our supLM tests is valid for the innovation process being a martingale difference sequence but does not take into account GARCH-type innovations. A possible solution would be to adopt a wild-bootstrap scheme similar to that used in \cite{Cha20}. While the implementation is straightforward, to the best of our knowledge, the validity of the bootstrap in a threshold framework has not been proven, even for TAR models, and constitutes and interesting challenge for future investigations.
The analysis of the tree-ring time series shows that TARMA models can provide a new insight into all those problems that make use of dendrochronological data. The TARMA(1,1) fit improves considerably over the commonly accepted linear specification that did not account for a short term non-linear effect.

\section*{Supplementary Material}
The supplemental document contains additional results from both the simulation study and the tree-ring data analysis.


\appendix
\section{Proofs}\label{sec:proofs}
\subsection*{Proof of Lemma~\ref{lemma:score}}
We proceed by induction. For the sake of presentation, we detail below a specific case since the argument can be easily adapted to the general setting. Hence, take $p=q=2$ and we prove that
    \begin{equation}\label{eqn:score1}
      \frac{\partial\eps_t}{\partial\phi_{10}}=-\sum_{j=0}^{t-1}\alpha_j,\quad\text{ where }
    \quad\begin{cases}
    \alpha_j=0 & \text{if } j<0 \\
    \alpha_j=1 & \text{if } j=0\\
    \alpha_j=\theta_{11}\alpha_{j-1}+\theta_{12}\alpha_{j-2} & \text{if } j>0.
    \end{cases}
    \end{equation}
As $\eps_0=\eps_{-1}=0$ and $\partial\eps_t/\partial\phi_{10}=-1+\theta_{11}\partial\eps_{t-1}/\partial\phi_{10}+\theta_{12}\partial\eps_{t-2}/\partial\phi_{10}$, the result \ref{eqn:score1} is proved if we show that,  under the induction hypothesis, $\partial\eps_{t+1}/\partial\phi_{10}=-\sum_{j=0}^{t}\alpha_j$, which holds upon noting that:
\begin{align*}
\frac{\partial\eps_{t+1}}{\partial\phi_{10}}&=-\alpha_0-\left(\sum_{j=0}^{t-1}\theta_{11}\alpha_j+\sum_{j=-1}^{t-2}\theta_{12}\alpha_j\right)=-\alpha_0-\left[\sum_{j=0}^{t-1}\left(\theta_{11}\alpha_j+\theta_{12}\alpha_{j-1}\right)\right]\\
&=-\alpha_0-\sum_{j=1}^{t}\alpha_j=-\sum_{j=0}^{t}\alpha_j.
\end{align*}
The same argument holds for the other components and, hence, the proof is completed.

\subsection*{Proof of Lemma~\ref{lemma:Fisher}}
Let $\boldsymbol\lambda=(\boldsymbol\zeta^\intercal,\bPsi^\intercal)^\intercal$ and consider:
\begin{align*}
-\frac{\partial^2\ell}{\partial\lambda_{i}\partial\lambda_{j}}&=\frac{1}{\sigma^2}\sum_{t=1}^{n}\frac{\partial\eps_t}{\partial\lambda_{i}} \frac{\partial\eps_t}{\partial\lambda_{j}}+\frac{1}{\sigma^2}\sum_{t=1}^{n}\eps_t\frac{\partial^2\eps_t}{\partial\lambda_{i}\partial\lambda_{j}},
\end{align*}
where $\lambda_i$,$\lambda_j$ are elements of $\boldsymbol\lambda$. The claim will follow if we prove
$$\frac{1}{\sigma^2}\sum_{t=1}^{n}\eps_t\frac{\partial^2\eps_t}{\partial\lambda_{i}\partial\lambda_{j}}=o_p(n)\text{ uniformly on $r$.}$$
From Lemma~\ref{lemma:score} it follows that if $\lambda_{i}\neq \theta_{1s}$, $\lambda_{i}\neq \Psi_{2s}$, $\lambda_{j}\neq \theta_{1s}$ and $\lambda_{j}\neq\Psi_{2s}$, with $s\in\{1,\dots,q\}$ then $\partial^2\eps_t/(\partial\lambda_{i}\partial\lambda_{j})=0$. As for the other instances, we detail below one case since identical arguments work for the remaining cases. Consider  $\partial^2\eps_t/(\partial\phi_{11}\partial\theta_{11})$.
The following recursive formula holds:
 \begin{align*}
 \frac{\partial^2\eps_t}{\partial\phi_{11} \partial\theta_{11}}&=\frac{\partial\eps_{t-1}}{\partial\phi_{11}} + \sum_{s=1}^{q}\theta_{1s}\frac{\partial^2\eps_{t-s}}{\partial\phi_{11}\partial\theta_{11}}
 =\sum_{k=0}^{t-1}\alpha_k\frac{\partial\eps_{t-1-k}}{\partial\phi_{11}}
 =\sum_{k=0}^{t-1}\alpha_k\sum_{j=0}^{t-2-k}\alpha_jX_{t-2-k-j}.
 \end{align*}
 The proof is completed if we show that
 \begin{align*}
 \frac{1}{n}\frac{1}{\sigma^2}\sum_{t=1}^{n}\eps_t\frac{\partial^2\eps_t}{\partial
 \phi_{11}\partial\theta_{11}}
 &=\frac{1}{n\sigma^2}\sum_{t=1}^{n}\eps_t \frac{\partial\eps_{t-1}}{\partial\phi_{11}}+\frac{1}{n}\frac{1}{\sigma^2} \sum_{t=1}^{n}\eps_t\sum_{s=1}^{q}\theta_{1s} \frac{\partial^2\eps_{t-s}}{\partial\phi_{11}\partial\theta_{11}}=o_p(1)
 \end{align*}
 It is easy to see that $n^{-1}\sum_{t=1}^{n}\eps_t(\partial\eps_{t-1})/(\partial\phi_{11})=o_p(1)$. Indeed, it is a martingale difference sequence whose variance equals:
 \begin{align*}
 V\left(\frac{1}{n}\sum_{t=1}^{n}\eps_t\frac{\partial\eps_{t-1}}{\partial\phi_{11}}\right)\leq&\frac{K}{n^2}\sum_{t=1}^{n}\sum_{j=0}^{t-1}|\alpha_j|E[X_{t-1-j}^2],
 \end{align*}
 which converges to zero as $n$ increases. Here $K$ is a constant that depends only on $\theta$'s and $\sigma^2$; we can take $K=\sigma^2\sum_{j=0}^{\infty}|\alpha_j|$.
 Similarly, we have that $n^{-1}\sum_{t=1}^{n}\eps_t(\theta_s\partial^2\eps_{t-s})/(\partial\phi_{11}\partial\theta_{1s})$
 is $o_p(1)$ for each $s\in\{1,\dots,q\}$ so that the proof is completed.
\subsection*{Proof of Proposition~\ref{prop:asymptotics}}

\begin{description}
\item[Part 1.] The proof for $\sLMg$ can be found in \cite{Li11} so that we do not repeat it here. In the following we prove the proposition for the $\sLM$ statistic, by showing that
\[
\Lambda(r)=
\begin{pmatrix}
\Lambda_{11} & \Lambda_{12}(r) \\
\Lambda_{12}^\intercal(r) & \Lambda_{22}(r)
\end{pmatrix}
= E\left[\frac{\partial\varepsilon_t}{\partial\boldsymbol\lambda} \left(\frac{\partial\varepsilon_t}{\partial\boldsymbol\lambda}\right)^\intercal\right]
\]
is positive definite, where $\boldsymbol{\lambda}=(\boldsymbol\phi^\intercal,\boldsymbol\Psi^\intercal)\in\mathds{R}^{2(p+1)}$
Since the matrix is symmetric it is sufficient to show that if
\begin{equation}\label{eqn:quadratic_form}
  E\left[c^\intercal\frac{\partial\varepsilon_t}{\partial\boldsymbol\lambda}\left(\frac{\partial\varepsilon_t}{\partial\boldsymbol\lambda}\right)^\intercal c\right]=0
\end{equation}
then $c$ is the $2(p+1)$-length zero vector. Equation~\ref{eqn:quadratic_form} holds if and only if
$
c^\intercal\frac{\partial\varepsilon_t}{\partial\boldsymbol\lambda}=0\text{ a.s.}.
$
Hereafter, $c=(c_{10},c_{11},\ldots,c_{1p},c_{20},c_{21},\ldots,c_{2p})^\intercal.$
 Routine algebra implies that
\begin{align}
\left[c_{10}+\sum_{k=1}^{p}c_{1k}X_{t-k}\right]I(X_{t-d}> r)=0\text{ a.s.}\label{eqn:form_quad1}\\
\left[(c_{10}+c_{20})+\sum_{k=1}^{p}(c_{1k}+c_{2k})X_{t-k}\right]I(X_{t-d}\leq r)=0\text{ a.s.}\label{eqn:form_quad2}
\end{align}
We proceed by contradiction: we assume that $\mathbf{c}$ is not the zero vector and prove that equalities (\ref{eqn:form_quad1}) and (\ref{eqn:form_quad2}) do not hold. Let $\mathcal{C}=\left\{c_{10}+\sum_{k=1}^{p}c_{1k}X_{t-k}=0\right\}$. Since $\mathbf{c}\neq \boldsymbol 0$, at least one of its components is different from zero. Let $\iota=\min\{1\leq k\leq p: c_{1k}\neq0\}$. Below, we detail the proof for the case $\iota=1$ and then show how to modify the argument when $\iota>1$. For simplicity, assume  $c_{11}=1$. Hence, under $H_0$,  $\mathcal{C}=\left\{\varepsilon_{t-1}= \Upsilon_{t-2}\right\}$
with
$$\Upsilon_{t-2}=\sum_{s=1}^{q}\theta_{1s} \varepsilon_{t-1-s}-\phi_{10} -\sum_{k=1}^{p} \phi_{1k} X_{t-1-k}-c_{10} - \sum_{k=2}^{p}c_{1k}X_{t-k}.$$
$\Upsilon_{t-2}$ belongs to the sigma-algebra $\mathfrak{I}_{t-2}$ generated by $\varepsilon_{t-2},\varepsilon_{t-3},\ldots$ and, therefore, since $\varepsilon_{t-1}$ is independent of $\Upsilon_{t-2}$ and it has a density function, the law of iterated expectations implies that
$$P(\mathcal{C})=E[I(\varepsilon_{t-1}=\Upsilon_{t-2})]=E[E[I(\varepsilon_{t-1}=\Upsilon_{t-2})|\mathfrak{I}_{t-2}]]=0.$$
Since in $\mathcal{C}^c$ it holds that $\left[c_{10}+\sum_{k=1}^{p}c_{1k}X_{t-k}\right]\neq 0$ and $P(\mathcal{C}^c)=1$, we have
\begin{align*}
P\left(\left\{\left[c_{10}+\sum_{k=1}^{p}c_{1k}X_{t-k}\right]I(X_{t-d}> r)=0\right\}\right)
=&P\left(\left\{X_{t-d}\leq r\right\}|\mathcal{C}^c\right),
\end{align*}
which is positive because the density of $\varepsilon_t$ is positive everywhere implying that the stationary distribution of $\{X_t\}$ is positive everywhere.
This contradicts equality (\ref{eqn:form_quad1}) and hence $c_{11}$ must be zero. If $\iota>1$, the event $\mathcal{C}$ reduces to $\{\varepsilon_{t-\iota}=\Upsilon_{t-\iota-1}\}$, with $\Upsilon_{t-\iota-1}\in\mathfrak{I}_{t-\iota-1}$ therefore the same argument shows that $c_{10}=c_{12}=\ldots=c_{1p}=0$. Lastly, by using (\ref{eqn:form_quad2}) instead of (\ref{eqn:form_quad1}), we prove that $c_{20}=c_{21}=\ldots=c_{2p}=0$ and this completes the proof.
\item[Part 2.] In this case the proofs are the same for the two supLM statistics. The result follows if we prove that
   \begin{align*}
   &\sup_{r\in[a,b]}\left\|M_n(r)\right\|=o_p(1), \quad \text{where} \quad M_n(r)= \frac{\mathcal{I}_n(r)}{n}-\Lambda(r)
\end{align*}
To begin with, we prove that, for any fixed $r$, $M_n(r)\to 0$ in probability entrywise. For the sake of presentation, we detail the proof for a specific case since it can be easily adapted to the other entries. Consider the $(i_1,i_2)$-th component of $n^{-1}\mathcal{I}_n(r)$ with $i_1i_2\neq0$:
$$-\frac{1}{\sigma^2}\frac{1}{n}\sum_{t=1}^{n}\left\{\sum_{j_1=0}^{t-1}\alpha_{j_1}X_{t-i_1-j_1}I_r(X_{t-d-j_1})\right\} \left\{\sum_{j_2=0}^{t-1}\alpha_{j_2}X_{t-i_2-j_2} I_r(X_{t-d-j_2})\right\}.$$
It is not hard to prove that
\begin{align*}
&\left|\frac{1}{\sigma^2}\frac{1}{n}\sum_{t=1}^{n}\left\{\sum_{j_1=0}^{\infty}\alpha_{j_1}X_{t-i_1-j_1}I_r(X_{t-d-j_1})\right\} \left\{\sum_{j_2=0}^{\infty}\alpha_{j_2}X_{t-i_2-j_2} I_r(X_{t-d-j_2})\right\}\right.\\
&\left.-\frac{1}{\sigma^2}\frac{1}{n}\sum_{t=1}^{n}\left\{\sum_{j_1=0}^{t-1}\alpha_{j_1}X_{t-i_1-j_1}I_r(X_{t-d-j_1})\right\} \left\{\sum_{j_2=0}^{t-1}\alpha_{j_2}X_{t-i_2-j_2} I_r(X_{t-d-j_2})\right\}\right|
\end{align*}
is a $o_p(1)$ uniformly on $r$. Hence, the ergodicity of $\{X_t\}$ implies that $M_n(r)$ converges to zero in probability for each $r$. Now, Fix $a<b$ and consider a grid $a=r_0<r_1<\ldots< r_m=b$ with equal mesh size, i.e. $r_i-r_{i-1}\equiv c$, for some $c>0$.
It holds that
$$\sup_{r \in [r_{i-1}, r_i]}\|M_n(r)-M_n(r_{i-1})\|\le C_n\quad\text{for all $i$}.$$
Moreover, $E(C_n)\to 0 $ as  $c\to 0$. Because for any $r \in [a,b]$, there exists an $i$ such that $r_{i-1} \le r \le r_{i}$ and hence
$$M_n(r)=M_n(r)- M_n(r_{i-1})+M_n(r_{i-1}) \text{ and }
\sup_{r\in [a,b]}\|M_n(r)\|\le \max_{i=0,\ldots,m} M_n(r_i)+C_n.$$
The proof is complete since for fixed $m$, $\max_{i=0,\ldots,m} M_n(r_i) \to 0 $ in probability and $E(C_n)\to 0$ as $c\to 0$ in probability.
\item[Part 3] For this part the proof for the two statistics are similar and we show that for $\sLMg$. For the sake of presentation and without loss of generality, we focus on the TARMA$(1,1)$ case. Within this proof, all the $o_p(1)$ terms hold uniformly on $r\in[a,b]$. We need to prove that:
\begin{align*} \sup_{r\in[a,b]}&\left\|\frac{1}{\sqrt{n}} \frac{\partial\hat\ell}{\partial\boldsymbol\Psi}(r)-\left(\nabla_{n,2}(r)-\Lambda_{21}(r)\Lambda_{11}^{-1}\nabla_{n,1}\right)\right\| = o_p(1).
\end{align*}
Since $\sqrt{n}(\boldsymbol{\hat\zeta}-\boldsymbol\zeta)=\Lambda_{11}^{-1}n^{-1/2}\partial\ell/\partial\boldsymbol\zeta+o_p(1)$ and $\Lambda_{21}(r)=O_p(1)$ uniformly in $r\in[a,b]$, then it is sufficient to prove
\begin{align}\label{eqn:mainApprox}
\sup_{r\in[a,b]} \left\|\frac{1}{\sqrt{n}}\frac{\partial\hat\ell}{\partial\boldsymbol\Psi}(r) - \frac{1}{\sqrt{n}}\frac{\partial\ell}{\partial\boldsymbol\Psi}(r)+\Lambda_{21}(r)\sqrt{n}(\boldsymbol{\hat\zeta}-\boldsymbol\zeta)\right\|
=o_p(1).
\end{align}
 We prove (\ref{eqn:mainApprox}) componentwise; below we detail the argument for the first component. Thence, we show that
\begin{align}
\sup_{r\in[a,b]}&\left\|\frac{1}{\sqrt{n}}\frac{\partial\hat\ell}{\partial\Psi_{10}}(r) - \frac{1}{\sqrt{n}}\frac{\partial\ell}{\partial\Psi_{10}}(r) + \Lambda_{21}(r)_{1,1}\sqrt{n}(\hat\phi_{10} - \phi_{10})\right.\\
&\left. + \Lambda_{21}(r)_{1,2}\sqrt{n}(\hat\phi_{11}-\phi_{11}) + \Lambda_{21}(r)_{1,3}\sqrt{n}(\hat\theta_{11}-\theta_{11})\right\|=o_p(1),\nonumber
\end{align}
with $\Lambda_{21}(r)_{i,j}$ being the $(i,j)$-th component of matrix  $\Lambda_{21}(r)$.
\par Routine algebra implies that:
\begin{equation}\label{c}
\varepsilon_t - \hat\varepsilon_t  =(\hat\phi_{10}-\phi_{10}) \sum_{j=0}^{t-1}\hat\theta_{11}^j+(\hat\phi_{11}-\phi_{11})\sum_{j=0}^{t-1}\hat\theta_{11}^jX_{t-1-j}
+(\theta_{11}-\hat\theta_{11})\sum_{j=0}^{t-1}\hat\theta_{11}^j\varepsilon_{t-1-j}
\end{equation}
and
\begin{align}
\frac{\partial\varepsilon_t}{\partial\Psi_{10}}-\frac{\partial\hat\varepsilon_t}{\partial\Psi_{10}}&=(\theta_{11}-\hat\theta_{11})\sum_{j=0}^{t-1}\hat\theta_{11}^j\frac{\partial\varepsilon_{t-1-j}}{\partial\Psi_{10}}.\label{d}
\end{align}
Since  $1/\hat\sigma^2-1/\sigma^2=O_p(n^{-1/2})$, by omitting a negligible additive term, it follows that
 \begin{align*}
 \frac{1}{\sqrt{n}}\frac{\partial\hat\ell}{\partial\Psi_{10}}= -\frac{1}{\sqrt{n}}\sum_{t=1}^{n}\frac{\hat\varepsilon_t}{\sigma^2} \frac{\partial\hat\varepsilon_t}{\partial\Psi_{10}}
 =\frac{1}{\sqrt{n}}\frac{\partial\ell}{\partial\Psi_{10}}+\frac{1}{\sqrt{n}}\sum_{t=1}^{n} \frac{\varepsilon_t}{\sigma^2}\frac{\partial\varepsilon_t}{\partial\Psi_{10}}-\frac{1}{\sqrt{n}} \sum_{t=1}^{n}\frac{\hat\varepsilon_t}{\sigma^2}\frac{\partial\hat\varepsilon_t}{\partial\Psi_{10}}.
 \end{align*}
 The results follow if we prove that:
 \begin{align}\label{:4.3}
 &\frac{1}{\sqrt{n}}\sum_{t=1}^{n}\frac{\varepsilon_t}{\sigma^2}\frac{\partial\varepsilon_t}{\partial\Psi_{10}}- \frac{1}{\sqrt{n}}\sum_{t=1}^{n}\frac{\hat\varepsilon_t}{\sigma^2}\frac{\partial\hat\varepsilon_t}{\partial\Psi_{10}}
+\frac{1}{\sigma^2}\frac{1}{n}\sqrt{n}(\hat\phi_{10}-\phi_{10})\sum_{t=1}^{n}\frac{\partial\varepsilon_t}{\partial\Psi_{10}} \frac{\partial\varepsilon_t}{\partial\phi_{10}} + \nonumber\\ &\frac{1}{\sigma^2}\frac{1}{n}\sqrt{n}(\hat\phi_{11}-\phi_{11})\sum_{t=1}^{n}\frac{\partial\varepsilon_t}{\partial\Psi_{10}} \frac{\partial\varepsilon_t}{\partial\phi_{11}}
 +{\frac{1}{\sigma^2}\frac{1}{n}\sqrt{n}(\hat\theta_{11}-\theta_{11}) \sum_{t=1}^{n}\frac{\partial\varepsilon_t}{\partial\Psi_{10}} \frac{\partial\varepsilon_t}{\partial\theta_{11}}} = o_p(1).
 \end{align}
 By using (\ref{c}) and (\ref{d}), we have that:
 \begin{align*}
 &\frac{1}{\sqrt{n}}\sum_{t=1}^{n}\frac{\varepsilon_t}{\sigma^2} \frac{\partial\varepsilon_t}{\partial\Psi_{10}}-\frac{1}{\sqrt{n}} \sum_{t=1}^{n}\frac{\hat\varepsilon_t}{\sigma^2}\frac{\partial\hat\varepsilon_t}{\partial\Psi_{10}}\\
 &=\frac{1}{\sqrt{n}\sigma^2}\sum_{t=1}^{n}\frac{\partial\varepsilon_t}{\partial\Psi_{10}}(\hat\phi_{10}-\phi_{10}) \sum_{j=0}^{t-1}\hat\theta_{11}^j+ \frac{1}{\sqrt{n}\sigma^2} \sum_{t=1}^{n} \frac{\partial\varepsilon_t}{\partial\Psi_{10}}(\hat\phi_{11}-\phi_{11}) \sum_{j=0}^{t-1}\hat\theta_{11}^jX_{t-1-j}\\
 &+\frac{1}{\sqrt{n}\sigma^2} \sum_{t=1}^{n}\frac{\partial\varepsilon_t}{\partial\Psi_{10}}(\theta_{11} -\hat\theta_{11})\sum_{j=1}^{t-1}\hat{\theta}_{11}^j\varepsilon_{t-1-j}
 +\frac{1}{\sqrt n}\sum_{t=1}^{n}\frac{\hat\varepsilon_t}{\sigma^2}(\theta_{11}-\hat\theta_{11}) \sum_{j=0}^{t-1}\hat\theta_{11}^j\frac{\partial\varepsilon_{t-1-j}}{\partial\Psi_{10}}.
 \end{align*}
 Hence Eq~(\ref{:4.3}) follows upon proving that
\begin{align}
&\sup_{r\in[a,b]}\left\|\frac{1}{\sigma^2}\frac{1}{n}\sqrt{n}(\hat\phi_{10}-\phi_{10})\sum_{t=1}^{n}\frac{\partial\varepsilon_t}{\partial\Psi_{10}}\frac{\partial\varepsilon_t}{\partial\phi_{10}}+\frac{1}{\sqrt{n}\sigma^2}\sum_{t=1}^{n}\frac{\partial\varepsilon_t}{\partial\Psi_{10}}(\hat\phi_{10}-\phi_{10})\sum_{j=0}^{t-1}\hat\theta_{11}^j\right\|=o_p(1),\label{:4.3.a}\\
&\sup_{r\in[a,b]}\left\|\frac{1}{\sigma^2}\frac{1}{n}\sqrt{n}(\hat\phi_{11}-\phi_{11})\sum_{t=1}^{n}\frac{\partial\varepsilon_t}{\partial\Psi_{10}}\frac{\partial\varepsilon_t}{\partial\phi_{11}}+\frac{1}{\sqrt{n}\sigma^2}\sum_{t=1}^{n}\frac{\partial\varepsilon_t}{\partial\Psi_{10}}(\hat\phi_{11}-\phi_{11})\sum_{j=0}^{t-1}\hat\theta_{11}^jX_{t-1-j}\right\|=o_p(1),\label{:4.3.b}\\
&\sup_{r\in[a,b]}\left\|\frac{1}{\sigma^2}\frac{1}{n}\sqrt{n}(\hat\theta_{11}-\theta_{11})\sum_{t=1}^{n}\frac{\partial\varepsilon_t}{\partial\Psi_{10}}\frac{\partial\varepsilon_t}{\partial\theta_{11}}+\frac{1}{\sqrt{n}\sigma^2}\sum_{t=1}^{n}\frac{\partial\varepsilon_t}{\partial\Psi_{10}}(\theta_{11}-\hat\theta_{11})\sum_{j=1}^{t-1}\hat{\theta}_{11}^j\varepsilon_{t-1-j}\right\|=o_p(1),\label{:4.3.c}\\
&\sup_{r\in[a,b]}\left\|\frac{1}{\sqrt n}\sum_{t=1}^{n}\frac{\hat\varepsilon_t}{\sigma^2}(\theta_{11}-\hat\theta_{11})\sum_{j=0}^{t-1}\hat\theta_{11}^j\frac{\partial\varepsilon_{t-1-j}}{\partial\Psi_{10}}\right\|=o_p(1).\label{:4.3.d}
\end{align}
In the following, we prove Eq~\ref{:4.3.a}.
 \begin{align*}
 &\frac{1}{\sigma^2}\frac{1}{n}\sqrt{n}(\hat\phi_{10}-\phi_{10})\sum_{t=1}^{n}\frac{\partial\varepsilon_t}{\partial\Psi_{10}}\frac{\partial\varepsilon_t}{\partial\phi_{10}}+\frac{1}{\sqrt{n}\sigma^2}\sum_{t=1}^{n}\frac{\partial\varepsilon_t}{\partial\Psi_{10}}(\hat\phi_{10}-\phi_{10})\sum_{j=0}^{t-1}\hat\theta_{11}^j\\
 =&\frac{1}{\sigma^2}\frac{1}{\sqrt{n}}(\hat\phi_{10}-\phi_{10})\sum_{t=1}^{n}\frac{\partial\varepsilon_t}{\partial\Psi_{10}}\left\{\sum_{j=0}^{t-1}\hat\theta_{11}^j-\sum_{j=0}^{t-1}\theta_{11}^j\right\}
 \end{align*}
 Because $\hat{\theta}_{11}$ is consistent and the true value $|\theta_{11}|<1$, there exists $0<\gamma<1$ such that the event $\mathcal{E}_n=\{ |\theta_{11}| <\gamma, |\hat{\theta}_{11}|<\gamma\}$ holds with  probability approaching 1  as $n\to\infty$. Thus, with no loss of generality,  $\mathcal{E}_n$  is assumed to hold. Consequently, there exists  a positive constant $K$ such that for all $t\ge 1$,
\begin{equation}\label{eqn:theta_approx}
\left|\sum_{j=0}^t \hat{\theta}_{11}^j -\sum_{j=0}^t \theta_{11}^j\right| \le |\hat{\theta}_{11} -\theta_{11}| \sum_{j=1}^t j \gamma^{j-1} \le K |\hat{\theta}_{11} -\theta_{11}|.
\end{equation}

Hence it follows that:
\begin{align*}
&\sup_{r\in[a,b]}\left|\frac{1}{\sqrt{n}\sigma^2}\sum_{t=1}^{n}\frac{\partial\varepsilon_t}{\partial\Psi_{10}}(\hat\phi_{10}-\phi_{10})\sum_{j=0}^{t-1}\hat\theta_{11}^j+\frac{1}{\sigma^2}\frac{1}{n}\sqrt{n}(\hat\phi_{10}-\phi_{10})\sum_{t=1}^{n}\frac{\partial\varepsilon_t}{\partial\Psi_{10}}\frac{\partial\varepsilon_t}{\partial\phi_{10}}\right|\\
&\leq\frac{K|\hat\phi_{10}-\phi_{10}||\hat\theta_{11}-\theta_{11}|}{\sqrt{n}\sigma^2}\sum_{t=1}^{n}\left|\theta_{11}^j\right|=\frac{K|\hat\phi_{10}-\phi_{10}||\hat\theta_{11}-\theta_{11}|}{\sqrt{n}\sigma^2(1-\gamma)}=o_p(1).
\end{align*}
By using the same argument we can handle Eq~\ref{:4.3.b} and Eq~\ref{:4.3.c}. Hence it remains to prove Eq~\ref{:4.3.d}. In the following we use $K$ to refer to a generic constant that can change across lines. Since $\theta_{11}-\hat\theta_{11}=O(n^{-1/2})$, we have
\begin{align*}
&\frac{1}{\sqrt n}\sum_{t=1}^{n}\frac{\hat\varepsilon_t}{\sigma^2}(\theta_{11}-\hat\theta_{11})\sum_{j=0}^{t-1}\hat\theta_{11}^j \frac{\partial\varepsilon_{t-1-j}}{\partial\Psi_{10}}\\
&=\frac{K}{n}\sum_{t=1}^{n}\hat\varepsilon_t\sum_{j=0}^{t-1}\theta_{11}^j\frac{\partial\varepsilon_{t-1-j}}{\partial\Psi_{10}}+\frac{K}{n} \sum_{t=1}^{n}\hat\varepsilon_t\sum_{j=0}^{t-1}(\hat\theta_{11}^j-\theta_{11}^j)\frac{\partial\varepsilon_{t-1-j}}{\partial\Psi_{10}}:= A_n+B_n.
\end{align*}
We prove separately that both $A_n$ and $B_n$ are $o_p(1)$. The former follows upon noting that $E[A_n]=0$ and
\begin{align*}
&\frac{1}{n^2}\sum_{t=1}^{n}E\left[\left(\hat\varepsilon_t\sum_{j=0}^{t-1}\theta_{11}^j\frac{\partial\varepsilon_{t-1-j}}{\partial\Psi_{10}}\right)^2\right] 
=\frac{K}{n^2}\sum_{t=1}^{n}E\left[\hat\varepsilon_t^2\right]\xrightarrow[n\to\infty]{}0,
\end{align*}
Similarly, the asymptotic negligibility of $B_n$ follows by using (\ref{eqn:theta_approx}) and the fact that $\partial\varepsilon_{t-1-j}/\partial\Psi_{10}$ is uniformly bounded by $(1-\gamma)^{-1}$ in magnitude. Hence, Eq~(\ref{:4.3.b}) is verified and the whole proof is completed.
\end{description}
\subsection*{Proof of Theorem~\ref{th:null}}
 Proposition~\ref{prop:asymptotics} implies that:
 \begin{align*}
 \sup_{r\in[a,b]}&\left\|T_n(r)-\nabla_{n}(r)^\intercal\left(-\Lambda_{21}(r)\Lambda_{11}^{-1},{I}_{m\times m}\right)^\intercal\right.\\
 &\left.\left(\Lambda_{22}(r)-\Lambda_{21}(r)\Lambda_{11}^{-1}\Lambda_{12}(r)\right)^{-1}\left(-\Lambda_{21}(r)\Lambda_{11}^{-1},{I}_{m\times m}\right)\nabla_{n}(r)\right\|=o_p(1),
 \end{align*}
where ${I}_{m\times m}$ is the ${m\times m}$ identity matrix, $m=p+1$ for the $\sLM$ statistic and  $m=p+q+1$ for the $\sLMg$ statistic. The result follows by applying the same proof of Theorem 2.1 in \cite{Lin05} and Theorem 1 of \cite{Li11}. The proof is the same for both supLM statistics.
\subsection*{Proof of Proposition~\ref{prop:lp}}
Under assumptions A1--A3, the proof for the two supLM statistics is similar to that in \cite{Li11} and is omitted.
Here we show the proof for the $\sLM$ statistic without assumption A3.
\begin{align*}
\text{under $H_{0,n}$:}&\quad \eps_t=\sum_{j=0}^{t-1}\alpha_j\triangle_{\boldsymbol\phi}(X_{t-1-j})-\phi_0\sum_{j=0}^{t-1}\alpha_j\\
\text{under $H_{1,n}$:}&\quad \eps_t=\sum_{j=0}^{t-1}\alpha_j\triangle_{\boldsymbol\phi}(X_{t-1-j})-\phi_0\sum_{j=0}^{t-1}\alpha_j-\sum_{j=0}^{t-1}\alpha_jh_n(X_{t-1-j}),
\end{align*}
where $\triangle_{\boldsymbol\phi}(X_t) = X_{t+1}-\sum_{k=1}^{p}\phi_{1k}X_{t-k}$ and
$$h_n(X_t)=\frac{1}{\sqrt{n}}\left(h_{10}+\sum_{k=1}^{p}h_{1k}X_{t-k}-h_{21}\sum_{s=1}^{q}\eps_{t-s}\right)I_{r_0}(X_{t-d}).$$
 Under the null hypothesis, we have that
 \begin{align*}
    \log\frac{d P_{1,n}}{d P_{0,n}}=&\mathbf{h}^\intercal\nabla_{n,2}(r_0)-\frac{1}{2}\mathbf{h}^\intercal \mathcal{I}_{n,22}(r_0)\mathbf{h}.
    \end{align*}
     Since $\mathcal{I}_{n,22}(r_0)$ converges to $\Lambda(r_0,r_0)$ and the martingale central limit theorem implies that $\nabla_{n,2}(r_0)$ converges to $\nabla_2(r_0)$, the proof of point $(i)$ is completed. To prove point $(ii)$ note that  such limiting random variable is Gaussian distributed with mean $-(\mathbf{h}^\intercal\Lambda(r_0,r_0)\mathbf{h})/2$ and variance $\mathbf{h}^\intercal\Lambda(r_0,r_0)\mathbf{h}$. The contiguity readily follows by applying Le Cam's first lemma (see example 6.5 pag 89 of \citet{Vaa98} concerning the asymptotic log normality of the log-likelihood ratio).
\subsection*{Proof of Theorem~\ref{th:loc_distr}}
Under assumptions A1--A3, the proof for the two supLM statistics is similar to that in \cite{Li11} and is omitted.
Here we show the proof for the $\sLM$ statistic without assumption A3. In order to prove point $(i)$ note that the tightness of $\nabla_{n,2}(r)-\Lambda_{21}(r)\Lambda_{11}\nabla_{n,1}$ under $H_{1,n}$ follows from its tightness under $H_{0,n}$ due to the contiguity. Hence, it suffices to show that, under $H_{1,n}$, $\nabla_{n,2}(r)-\Lambda_{21}(r)\Lambda_{11}^{-1}\nabla_{n,1}$ converges to $(\xi_r+\gamma_r)$ when $r$ is fixed. By virtue of the third Le Cam's Lemma, it suffices to prove that
     $$ \left(\nabla_{n,2}(r)-\Lambda_{21}(r)\Lambda_{11}^{-1}\nabla_{n,1},\log\frac{d P_{1,n}}{d P_{0,n}}\right)$$
    converges to a multivariate normal distributed random vector whose variance-covariance matrix is
    $$
    \begin{pmatrix}
    \Lambda(r,r)-\Lambda(r,\infty)\Lambda^{-1}(\infty,\infty)\Lambda(\infty,r) & \gamma_r \\
    \gamma_r^\intercal & \mathbf{h}^\intercal\Lambda(r_0,r_0)\mathbf{h}
    \end{pmatrix}
    $$
    which is true because:
    \begin{align*}
    \text{Cov}&\left(\nabla_{n,2}(r)-\Lambda_{21}(r)\Lambda_{11}^{-1}\nabla_{n,1},\log\frac{d P_{1,n}}{d P_{0,n}}\right)=\left\{\Lambda_{22}(\min\{r,r_0\})-\Lambda_{21}(r)\Lambda_{11}^{-1}\Lambda_{12}(r_0)\right\}\mathbf{h}.
    \end{align*}
The proof of point $(ii)$ is omitted since it follows directly from $(i)$.
\subsection*{Proof of Theorem~\ref{th:power}}
The proof is the same for both statistics. To prove the theorem, it suffices to show that for each $q$
\begin{equation}\label{eqn:power_final}
\Pr\left((\xi_{r_0}+\gamma_{r_0})^\intercal\left(\Lambda_{22}({r_0}) - \Lambda_{21}({r_0})\Lambda^{-1}_{11}\Lambda_{12}({r_0})\right)^{-1}(\xi_{r_0}+\gamma_{r_0})>q\right)\xrightarrow[\|\mathbf{h}\|\to+\infty]{p}1.
\end{equation}
For simplicity, we set $\Sigma_{r_0}=\left(\Lambda_{22}({r_0})-\Lambda_{21}({r_0})\Lambda^{-1}_{11}\Lambda_{12}({r_0})\right)$. Routine algebra implies that:
\begin{align*}
&(\xi_{r_0}+\gamma_{r_0})^\intercal\left(\Lambda_{22}({r_0})-\Lambda_{21}({r_0})\Lambda^{-1}_{11}\Lambda_{12}({r_0})\right)^{-1} (\xi_{r_0}+\gamma_{r_0})
=\xi_{r_0}^\intercal\Sigma_{r_0}^{-1}\xi_{r_0}+2\mathbf{h}^\intercal\xi_{r_0}+\mathbf{h}^\intercal\Sigma_{r_0}\mathbf{h},
\end{align*}
For any fixed $r_0$, $\xi_{r_0}$ is a Gaussian distributed random vector with zero mean and variance-covariance matrix $\Sigma_{r_0}$ and hence the Markov inequality implies that $\frac{\mathbf{h}^\intercal\xi_{r_0}}{\mathbf{h}^\intercal\Sigma_{r_0}\mathbf{h}}$ is $o_p(1)$ as $\|\mathbf{h}\|$ increases.
Therefore, we have
\begin{align*}
&\xi_{r_0}^\intercal\Sigma_{r_0}^{-1}\xi_{r_0}+2\mathbf{h}^\intercal\xi_{r_0}+\mathbf{h}^\intercal\Sigma_{r_0}\mathbf{h}
=\xi_{r_0}^\intercal\Sigma_{r_0}^{-1}\xi_{r_0}+2(\mathbf{h}^\intercal\Sigma_{r_0}\mathbf{h})\left(\frac{\mathbf{h}^\intercal\xi_{r_0}}{\mathbf{h}^\intercal\Sigma_{r_0}\mathbf{h}}+\frac{1}{2}\right)\\
&={O_p(1)+C_h(o_p(1)+1/2)\xrightarrow[\|\mathbf{h}\|\to+\infty]{p}+\infty}
\end{align*}
where $C_h=2(\mathbf{h}^\intercal\Sigma_{r_0}\mathbf{h})\to\infty$ as $\|\mathbf{h}\|\to+\infty$ and hence Eqn.~\ref{eqn:power_final} follows.

\bibliographystyle{apa}
\bibliography{ARMAvsTARMAv2}
\end{document}


\title{Supplement to: \\ Testing for threshold effects in the TARMA framework}

\author[2]{Greta Goracci}
\author[2]{Simone Giannerini}
\author[1]{Kung-Sik Chan}
\author[3,4,5]{Howell Tong}

\affil[1]{\small Department of Statistics and Actuarial Science,
University of Iowa, Iowa City, USA}
\affil[2]{Department of Statistical Sciences, University of Bologna, Italy}
\affil[3]{School of Mathematical Science, University of Electronic Science and Technology, Chengdu China}
\affil[4]{Center for Statistical Science, Tsinghua University, China}
\affil[5]{London School of Economics and Political Science, U.K.}

\date{\small \today}

\maketitle


\begin{abstract}
This Supplement has three sections. Section~\ref{sec:SMq} contains supplementary results for the tabulated quantiles of the null distribution, whereas Section~\ref{sec:SMMC} contains supplementary results from the simulation study. Lastly, Section~\ref{sec:SMtree} presents additional results on the tree-ring data analysis.
\end{abstract}

%
%
%

\section{Supplementary results for the tabulated quantiles of the null distribution}\label{sec:SMq}

\begin{table}
\caption{Tabulated quantiles for the asymptotic null distribution of the supLM statistics for the threshold range 25th-75th percentiles. The first two columns denote the AR and the MA orders, respectively.}\label{tab:q1SM}
\centering
\begin{tabular}{ccrrrrrrrr}
&& \multicolumn{4}{c}{$\sLM$} & \multicolumn{4}{c}{$\sLMg$} \\
      \cmidrule(lr){3-6}  \cmidrule(lr){7-10}
  AR & MA & 90\% & 95\% & 99\% & 99.9\% & 90\% & 95\% & 99\% & 99.9\% \\
      \cmidrule(lr){3-6}  \cmidrule(lr){7-10}
  1 & 1 &  9.61 & 11.37 & 15.19 & 20.38 & 11.64 & 13.44 & 17.42 & 22.83\\
  2 & 1 & 11.53 & 13.41 & 17.22 & 22.17 & 13.48 & 15.46 & 19.63 & 25.60\\
  3 & 1 & 13.74 & 15.71 & 19.98 & 25.04 & 15.59 & 17.61 & 21.91 & 27.98\\
  4 & 1 & 15.65 & 17.68 & 22.25 & 27.44 & 17.42 & 19.52 & 24.02 & 29.94\\
      \cmidrule(lr){3-6} \cmidrule(lr){7-10}
  1 & 2 &  9.64 & 11.47 & 15.50 & 20.25 & 13.69 & 15.57 & 19.67 & 25.07\\
  2 & 2 & 11.71 & 13.48 & 17.61 & 22.49 & 15.59 & 17.58 & 21.95 & 28.17\\
  3 & 2 & 13.46 & 15.35 & 19.33 & 25.06 & 17.13 & 19.18 & 23.54 & 29.78\\
  4 & 2 & 15.55 & 17.58 & 21.82 & 27.80 & 18.97 & 21.21 & 26.42 & 31.92\\
      \cmidrule(lr){3-6} \cmidrule(lr){7-10}
\end{tabular}
\end{table}

It is easy to see from Table~\ref{tab:q1SM} that the asymptotic distribution of the supLM statistics depends only upon the size of the parameter vector $\bPsi$.  For instance the 90\% percentile of the $\sLMg$ statistic for the ARMA$(2,1)$ (13.48) is the same (up to sampling fluctuations) as that for the ARMA$(1,2)$ (13.69) since in both cases there are 3 parameters tested. Moreover, such value matches the percentiles of the $\sLM$ statistic for the ARMA$(3,1)$ (13.74) and ARMA$(3,2)$ (13.46) since, also in this case, only the 3 autoregressive parameters are tested. This evidence suggests that a unique table could be adopted for both tests, for instance that of \cite{And03}.
\par
The simulations show that the supLM statistics do not depend upon a specific parameters' choice. In order to investigate further the similarity we simulate from an ARMA$(1,1)$ model with different parameterizations, some of which present the problem of quasi-cancellation of the AR and MA polynomials. The quantiles are presented in Table~\ref{tab:q2} and show that the rate of convergence towards the asymptotic distribution of the test statistics might depend upon the true value of the parameters so that some small discrepancies are expected in corner cases. Still, the differences do not affect the practical usability of the tests; in case of small sample sizes, a conservative approach could be selecting an ARMA model by means of a consistent order selection procedure. Then, one could simulate the asymptotic null by using the estimated model and sample size in use.
%
\begin{table}
\caption{Tabulated quantiles for the asymptotic null distribution of the supLM statistics as in Table~\ref{tab:q1SM} but for different parameterizations of the ARMA$(1,1)$ process.}\label{tab:q2}
\centering
\begin{tabular}{rrrrrrrrrr}
&& \multicolumn{4}{c}{$\sLM$} & \multicolumn{4}{c}{$\sLMg$} \\
      \cmidrule(lr){3-6}  \cmidrule(lr){7-10}
$\phi$ & $\theta$ & 90\% & 95\% & 99\% & 99.9\% & 90\% & 95\% & 99\% & 99.9\% \\
      \cmidrule(lr){3-6}  \cmidrule(lr){7-10}
 -0.60 & -0.80 &  9.54 & 11.28 & 14.68 & 19.58 & 11.68 & 13.49 & 17.38 & 22.12\\
  0.60 & -0.80 & 10.21 & 11.72 & 15.84 & 20.48 & 12.17 & 13.95 & 17.81 & 22.17\\
 -0.60 & -0.40 &  9.49 & 11.19 & 14.95 & 19.90 & 11.64 & 13.42 & 17.54 & 23.45\\
  0.60 & -0.40 &  9.90 & 11.44 & 14.79 & 19.84 & 11.67 & 13.63 & 17.49 & 21.96\\
 -0.60 &  0.00 &  9.51 & 11.24 & 14.93 & 19.64 & 11.52 & 13.41 & 17.21 & 23.37\\
  0.60 &  0.00 &  9.76 & 11.41 & 14.98 & 19.89 & 11.61 & 13.54 & 17.57 & 22.18\\
 -0.60 &  0.40 &  9.68 & 11.40 & 15.23 & 19.26 & 11.89 & 13.75 & 17.87 & 22.52\\
  0.60 &  0.40 &  9.69 & 11.29 & 14.93 & 20.32 & 11.69 & 13.57 & 17.66 & 22.85\\
 -0.60 &  0.80 & 10.02 & 11.77 & 15.30 & 19.80 & 12.54 & 14.44 & 18.02 & 23.05\\
  0.60 &  0.80 &  9.99 & 11.81 & 15.51 & 20.82 & 12.17 & 14.03 & 18.02 & 23.03\\
      \cmidrule(lr){3-6}  \cmidrule(lr){7-10}
\end{tabular}
\end{table}

\section{Supplementary Monte Carlo results}\label{sec:SMMC}
\subsection{Power: TARMA$(2,1)$ model}\label{appB}
We simulate from the following TARMA$(2,1)$ model where only the autoregressive parameters change across regimes:
\begin{equation}\label{TARMA21}
 X_t =  - 0.35 X_{t-1} - 0.45 X_{t-2} - \theta_{11} \eps_{t-1} + \left(\Psi_{10} + \Psi_{11} X_{t-1} + \Psi_{12} X_{t-2}\right)I(X_{t-1}\leq 0) + \eps_t.
\end{equation}
where $\Psi_{10}=\Psi_{11}=\Psi_{12} = \Psi$ and $\theta_{11}$ are as in  Table~\ref{tab:p4}, that presents the size-corrected power of the tests. The results are consistent with the TARMA$(1,1)$ case $i)$ of Table~3 of the main article, where only the autoregressive parameters change. Also in this case the $\sLM$ test is almost uniformly superior to the $\sLMg$ test, that, in turn is more powerful than the qLR in 60\% of the cases.

\begin{table}
\centering
\caption{Size-corrected power for the TARMA$(2,1)$ model of Eq.~(\ref{TARMA21}). Sample size $n=100,200,500$, $\alpha=5\%$.}\label{tab:p4}
\begin{tabular}{rrrrrrrrrrr}
&& \multicolumn{3}{c}{$n=100$}& \multicolumn{3}{c}{$n=200$}& \multicolumn{3}{c}{$n=500$}\\
    \cmidrule(lr){3-5} \cmidrule(lr){6-8} \cmidrule(lr){9-11}
$\Psi$ & $\theta_{11}$ & $\sLM$ & $\sLMg$ & qLR & $\sLM$ & $\sLMg$ & qLR & $\sLM$ & $\sLMg$ & qLR \\
    \cmidrule(lr){3-5} \cmidrule(lr){6-8} \cmidrule(lr){9-11}
  0.2 & -0.5 & 16.4 & 13.3 & 14.0 & 28.5 & 27.4 & 31.2 &  81.0 &  77.9 &  65.5 \\
  0.6 & -0.5 & 84.0 & 78.9 & 74.6 & 98.1 & 98.0 & 98.7 & 100.0 & 100.0 & 100.0 \\
  0.8 & -0.5 & 94.5 & 93.7 & 93.5 & 98.7 & 98.9 & 100.0& 100.0 & 100.0 & 100.0 \\
    \cmidrule(lr){3-5} \cmidrule(lr){6-8} \cmidrule(lr){9-11}
  0.2 &  0.0 & 10.7 &  8.3 & 10.0 & 24.8 & 19.5 & 19.0 &  53.4 & 46.2  &  48.6 \\
  0.6 &  0.0 & 63.1 & 60.6 & 60.3 & 94.0 & 92.8 & 91.6 &  99.8 & 99.8  & 100.0 \\
  0.8 &  0.0 & 88.5 & 87.7 & 84.1 & 99.1 & 99.1 & 99.5 &  99.9 & 99.9  & 100.0 \\
    \cmidrule(lr){3-5} \cmidrule(lr){6-8} \cmidrule(lr){9-11}
  0.2 &  0.5 & 15.4 & 13.7 &  7.5 & 28.7 & 25.2 & 23.4 &  62.9 & 57.4  &  51.9 \\
  0.6 &  0.5 & 81.6 & 76.0 & 44.9 & 98.7 & 98.3 & 97.3 & 100.0 & 100.0 & 100.0 \\
  0.8 &  0.5 & 94.1 & 92.8 & 79.5 & 99.8 & 99.8 & 99.9 & 100.0 & 100.0 & 100.0 \\
    \cmidrule(lr){3-5} \cmidrule(lr){6-8} \cmidrule(lr){9-11}
\end{tabular}
\end{table}

\subsection{The impact of model selection in presence of mis-specification}\label{appC}

In Table~\ref{tab:ms3} we show the results of the Hannan-Rissanen model selection in presence of modelling misspecification. As for the size, the bias produced is small and stays within 3\%. The case of power is more interesting. While there is little or no power loss, there might be a substantial gain for the $\sLM$ test, see e.g. the TAR3 instance, where the power for $n=100$ passes from 34.7\% to 96.1\%, most likely due to the HR criterion picking the dependence upon lag 3.

\begin{table}
\caption{Rejection percentages under model misspecification for the processes of Table~6 of the main article for the supLM tests ($\sLM$, $\sLMg$). The subscript ``\textsc{HR}'' indicates that the order of the ARMA model has been selected through the Hannan-Rissanen procedure, otherwise, an ARMA$(1,1)$ model is tested. The upper panel (linear processes) reflects the empirical size at nominal level $5\%$. The lower panel (non-linear processes) reflects the empirical power for non-linear processes that are not representable as a TARMA$(1,1)$ process.}\label{tab:ms3}
\centering 
\begin{tabular}{rrrrrrrr}
&& \multicolumn{2}{c}{$n=100$}& \multicolumn{2}{c}{$n=200$}& \multicolumn{2}{c}{$n=500$}\\
   \cmidrule(lr){3-4} \cmidrule(lr){5-6} \cmidrule(lr){7-8}
&& $\sLM$ & $\sLMh$ & $\sLM$ & $\sLMh$  & $\sLM$ & $\sLMh$  \\
   \cmidrule(lr){3-4} \cmidrule(lr){5-6} \cmidrule(lr){7-8}
\multirow{7}{*}{\rotatebox[origin=c]{90}{\textsc{linear}}}&
   AR5       & 7.9 &10.2 & 5.5 & 5.3 & 4.6 & 5.7 \\
&  AR2.1     & 6.7 & 7.7 & 4.7 & 4.9 & 5.3 & 5.0 \\
&  AR2.2     & 8.7 & 9.5 & 4.1 & 5.7 & 3.4 & 4.9 \\
&  ARMA21.1  & 5.9 & 7.2 & 5.5 & 6.3 & 5.5 & 6.2 \\
&  ARMA21.2  & 5.8 & 6.4 & 4.5 & 5.0 & 4.0 & 5.1 \\
&  ARMA22    & 5.5 & 6.2 & 3.7 & 3.7 & 3.5 & 3.7 \\
&  MA2       & 3.5 & 6.9 & 2.3 & 3.1 & 4.4 & 5.3 \\
   \cmidrule(lr){3-4} \cmidrule(lr){5-6} \cmidrule(lr){7-8}
\multirow{10}{*}{\rotatebox[origin=c]{90}{\textsc{non-linear}}}&
     TAR3    &  34.7 & 96.1 & 61.8 & 99.9 &  96.7 & 100.0 \\
&    3TAR1   &  19.2 & 19.5 & 36.1 & 35.9 &  78.5 &  78.1 \\
&    NLMA.1  &  85.1 & 84.7 & 97.3 & 97.4 &  99.8 &  99.8 \\
&    NLMA.2  &  86.4 & 84.5 & 98.3 & 98.6 & 100.0 & 100.0 \\
&    BIL.1   &  12.0 & 21.6 & 13.5 & 29.0 &  14.7 &  28.9 \\
&    BIL.2   &  84.0 & 82.0 & 98.7 & 97.9 & 100.0 &  99.8 \\
&    EXPAR.1 & 100.0 &100.0 &100.0 & 99.9 & 100.0 & 100.0 \\
&    EXPAR.2 &  95.8 & 97.7 & 99.6 & 99.9 & 100.0 & 100.0 \\
&    NLAR    & 100.0 &100.0 &100.0 &100.0 & 100.0 & 100.0 \\
   \cmidrule(lr){3-4} \cmidrule(lr){5-6} \cmidrule(lr){7-8}\\
&& \multicolumn{2}{c}{$n=100$}& \multicolumn{2}{c}{$n=200$}& \multicolumn{2}{c}{$n=500$}\\
   \cmidrule(lr){3-4} \cmidrule(lr){5-6} \cmidrule(lr){7-8}
&& $\sLMg$ & $\sLMgh$ & $\sLMg$ & $\sLMgh$  & $\sLMg$ & $\sLMgh$  \\
   \cmidrule(lr){3-4} \cmidrule(lr){5-6} \cmidrule(lr){7-8}
\multirow{7}{*}{\rotatebox[origin=c]{90}{\textsc{linear}}}&
   AR5       & 15.2 & 17.5 & 9.8 &10.5 & 6.9 & 7.2 \\
&  AR2.1     &  7.1 &  8.3 & 5.5 & 5.5 & 5.5 & 5.8 \\
&  AR2.2     & 10.1 &  9.9 & 5.8 & 5.5 & 5.0 & 5.7 \\
&  ARMA21.1  &  6.8 &  8.7 & 5.2 & 6.7 & 4.8 & 6.2 \\
&  ARMA21.2  &  8.9 &  9.8 & 5.7 & 6.0 & 4.9 & 5.6 \\
&  ARMA22    & 13.6 & 16.8 &12.7 &13.0 & 7.5 & 8.4 \\
&  MA2       & 12.1 & 14.9 & 5.8 & 7.2 & 6.2 & 8.1 \\
   \cmidrule(lr){3-4} \cmidrule(lr){5-6} \cmidrule(lr){7-8}
\multirow{10}{*}{\rotatebox[origin=c]{90}{\textsc{non-linear}}}&
     TAR3    & 98.1 & 99.6 & 99.7 &100.0 & 100.0 & 100.0 \\
&    3TAR1   & 18.4 & 18.8 & 30.4 & 29.9 &  72.4 &  71.5 \\
&    NLMA.1  & 84.0 & 83.6 & 97.2 & 96.9 &  99.9 &  99.9 \\
&    NLMA.2  & 86.8 & 84.2 & 98.5 & 98.3 &  99.9 &  99.9 \\
&    BIL.1   & 62.1 & 67.0 & 84.8 & 90.9 &  95.0 &  97.0 \\
&    BIL.2   & 83.1 & 81.5 & 98.9 & 97.8 & 100.0 &  99.9 \\
&    EXPAR.1 &100.0 &100.0 &100.0 & 99.8 & 100.0 & 100.0 \\
&    EXPAR.2 & 99.1 & 98.2 & 99.9 & 99.9 & 100.0 & 100.0 \\
&    NLAR    &100.0 &100.0 &100.0 &100.0 & 100.0 & 100.0 \\
   \cmidrule(lr){3-4} \cmidrule(lr){5-6} \cmidrule(lr){7-8}
\end{tabular}
\end{table}

\section{Supplementary results for the tree-ring data analysis}\label{sec:SMtree}
\begin{figure}
  \centering
  \includegraphics[width=.49\linewidth,keepaspectratio]{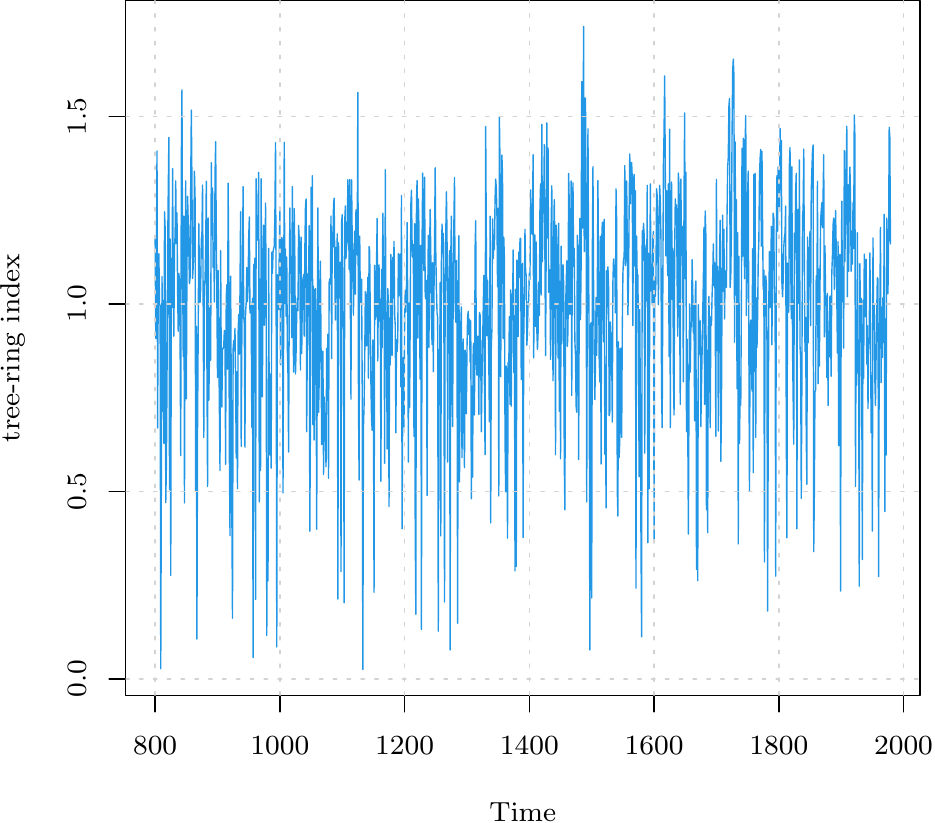}
  \includegraphics[width=.49\linewidth,keepaspectratio]{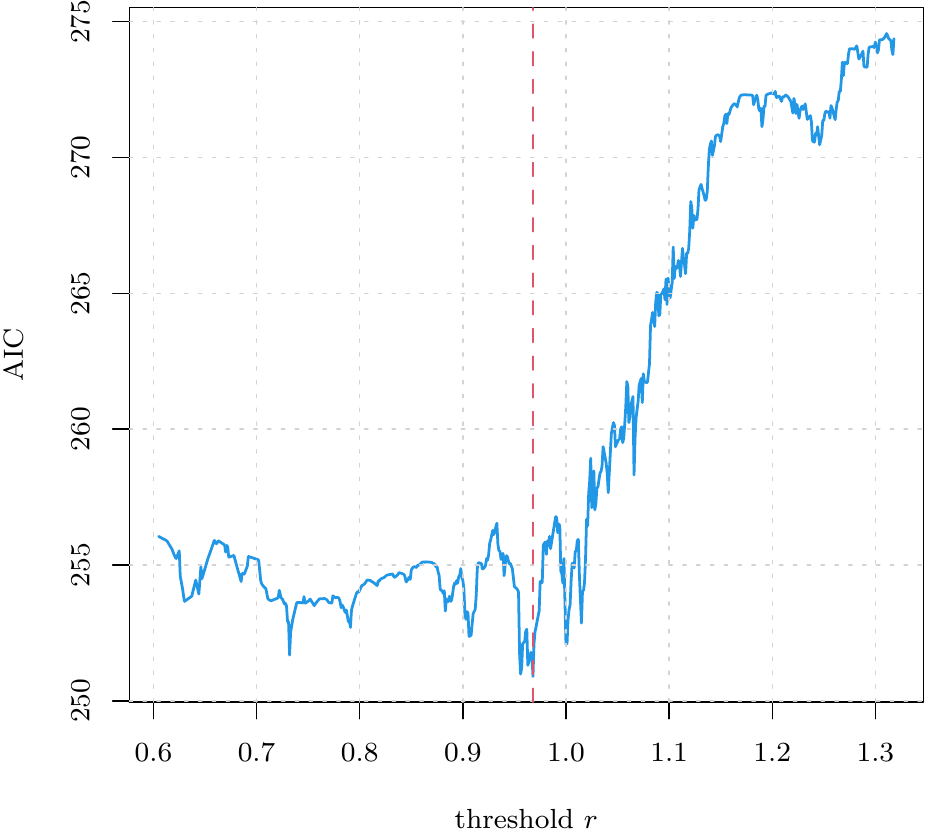}
 \caption{Time series of the yearly standardized tree-ring growth index of Pinus aristata var. longaeva, California, USA (left). AIC vs threshold $r$ (right). The value of $r$ that minimizes the AIC is indicated with a dashed red line.}\label{fig:tr1}
\end{figure}

\begin{figure}
  \centering
  \includegraphics[width=.9\linewidth,keepaspectratio]{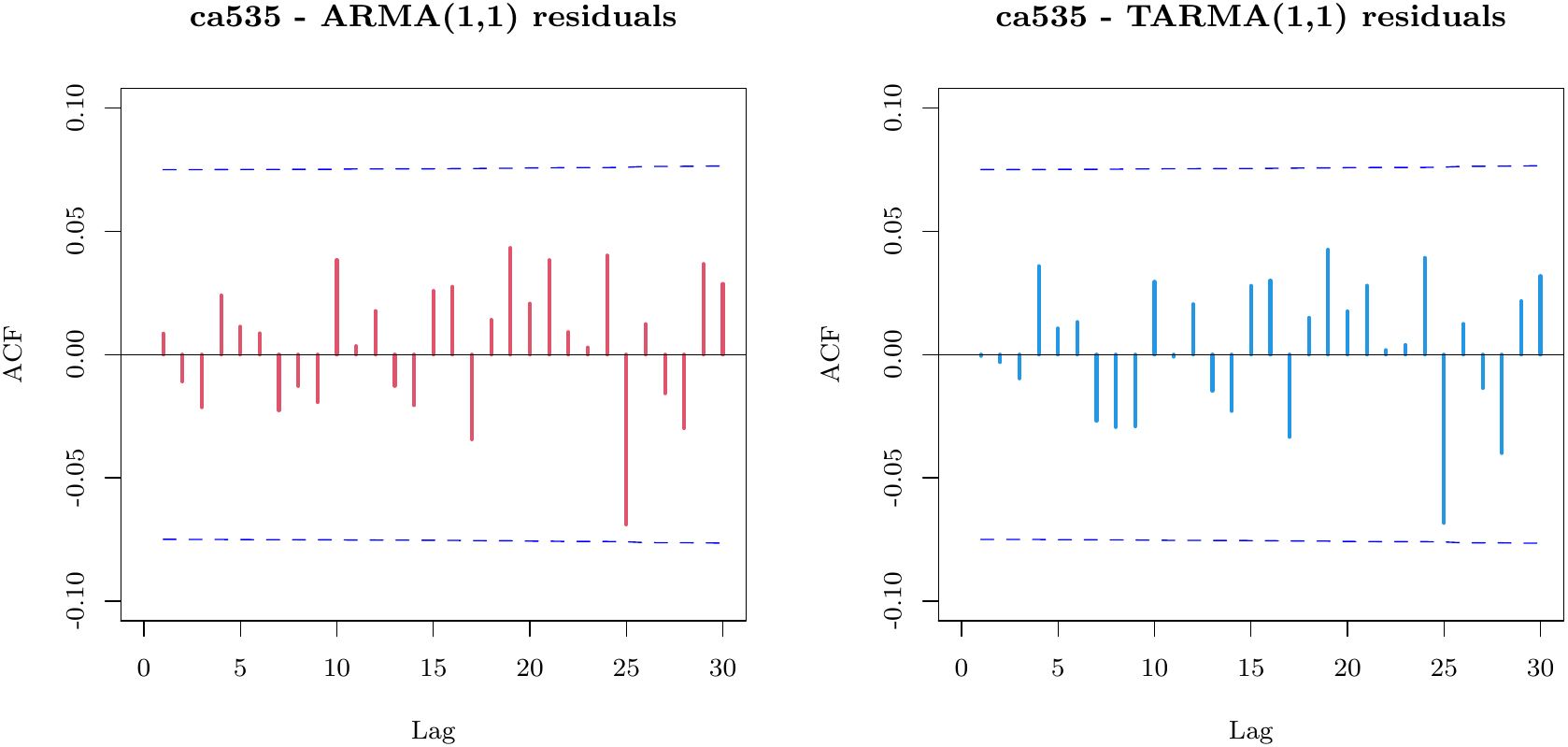}
  \includegraphics[width=.9\linewidth,keepaspectratio]{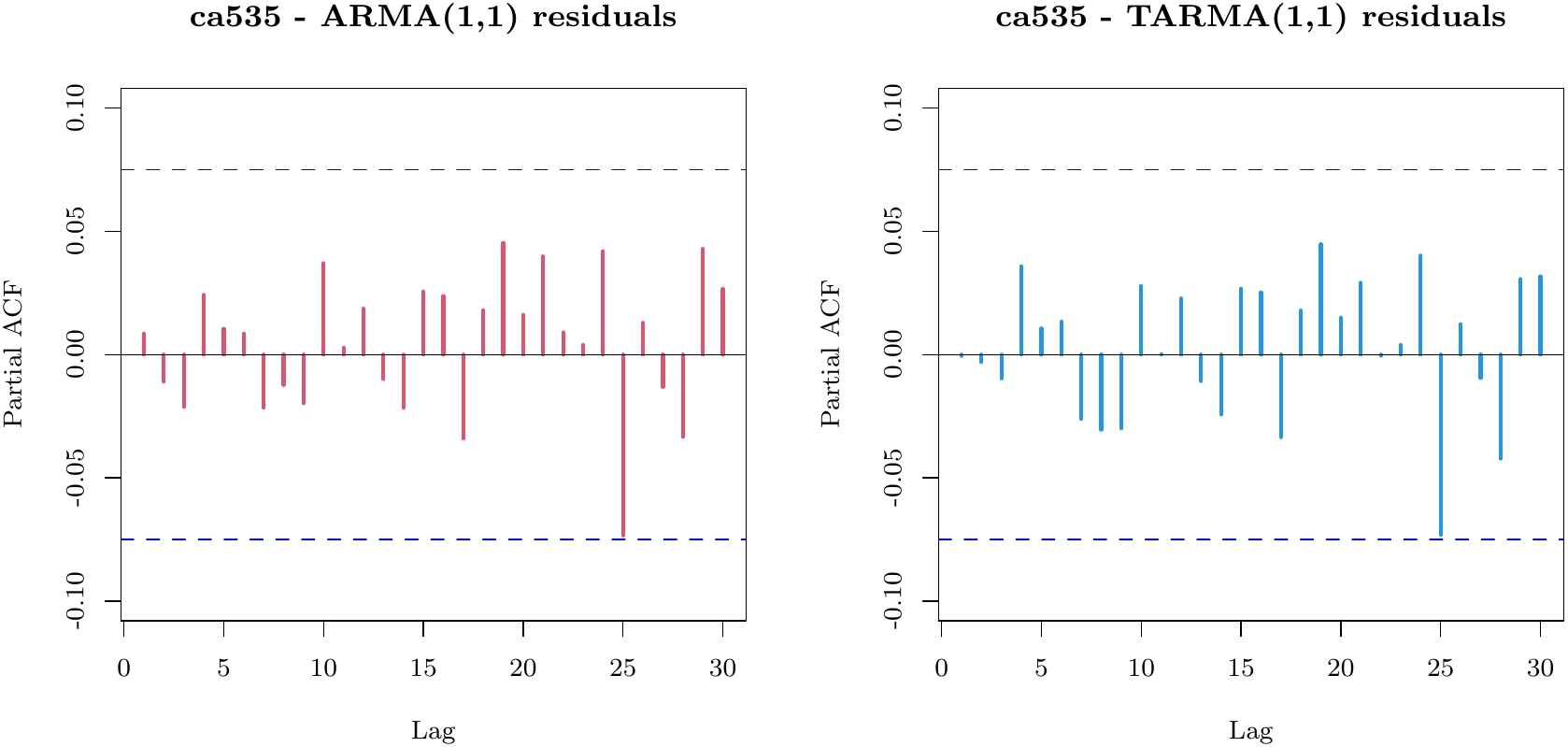}
  \caption{Global and partial correlograms up to 30 years computed over the residuals of the ARMA(1,1)(red/left) and TARMA(1,1)(blue/right) fits. The horizontal dashed lines correspond to the rejection bands of the null hypothesis of absence of correlation at level 99\%.}\label{fig:r.corr}
\end{figure}

\begin{figure}
  \centering
  \includegraphics[width=.9\linewidth,keepaspectratio]{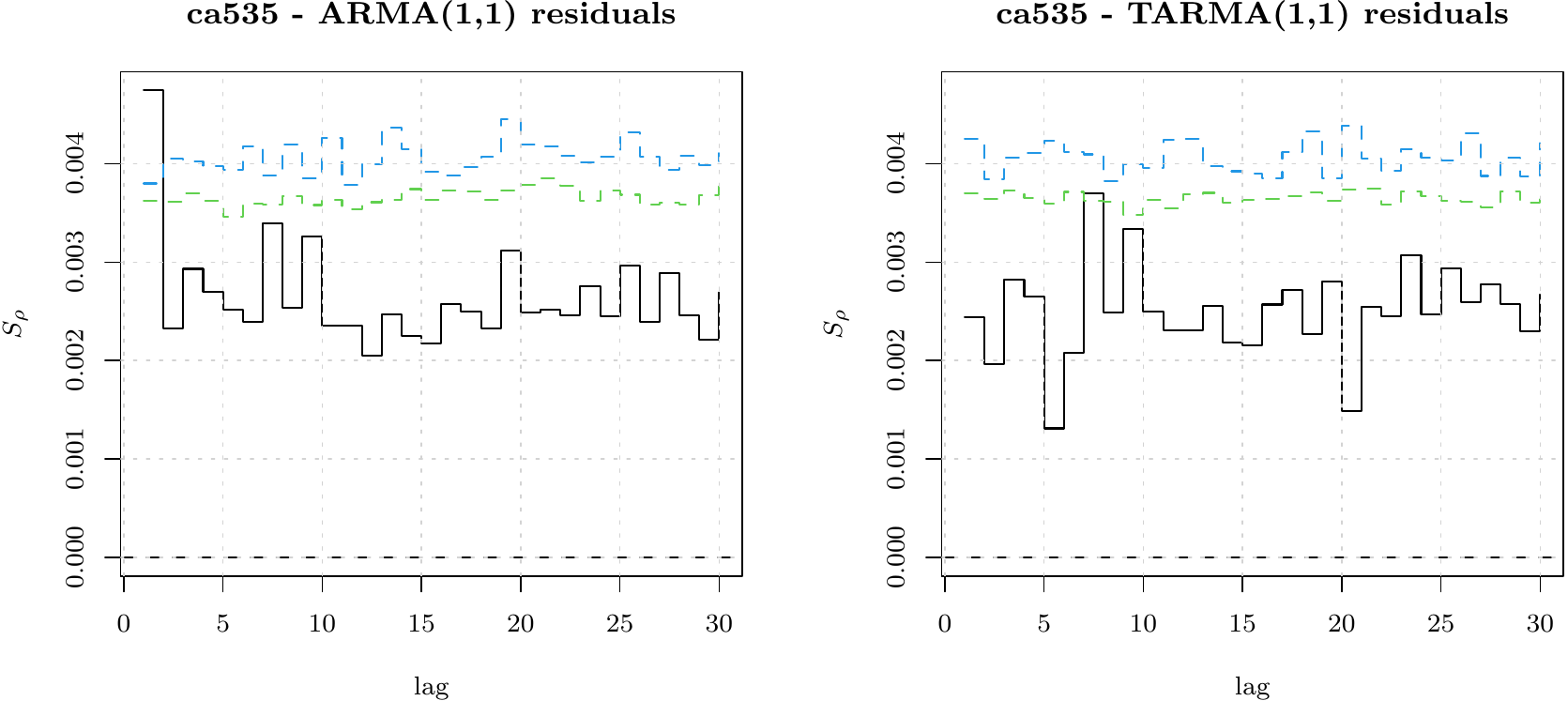}
  \caption{Entropy measure $S_{\rho}$ up to 30 years computed over the residuals of the ARMA(1,1) and TARMA(1,1) fits. The green and blue dashed lines correspond to the rejection bands of the null hypothesis of independence at level 99\% and 99.9\%, respectively.}\label{fig:r.srho}
\end{figure}

In Figure~\ref{fig:tr1}(left) we show the time series of the yearly standardized tree-ring growth index of  Pinus aristata var. longaeva, California, USA (left) (reconstructed by D.A. Graybill \cite{ca535}). The series ranges from year 800 to 1979 ($n=1180$). The vertical line in Figure~\ref{fig:tr1}(right) indicates the estimated threshold $\hat r = 0.97$, the same value of $r$ that maximizes the $T_n$ statistic.
The residual analysis is somehow revealing. The correlograms of the residuals for both the ARMA and TARMA fits of the are shown in Figure~\ref{fig:r.corr}. Clearly, there is no correlation structure left in the residuals of both fits and this may lead to conclude in favour of the ARMA(1,1) model. However, if the residuals are tested for independence with the entropy metric $S_{\rho}$ \citep{Gia15} then a different picture emerges. The results are shown in Figure~\ref{fig:r.srho}: the residuals of the ARMA$(1,1)$ fit (left panel)  present a significant unaccounted (non-linear) dependence at lag 1 whereas there is no dependence structure left in the residuals of the TARMA$(1,1)$ fit (right panel).

\bibliographystyle{elsarticle-harv}
\bibliography{ARMAvsTARMAv1}